\documentclass{aa}  

\usepackage{graphicx}
\usepackage{txfonts}
\usepackage{hyperref}
\hypersetup{
    colorlinks,
    linkcolor={blue!60!black},
    citecolor={blue!60!black},
    urlcolor={blue!60!black}
}

\usepackage{comment}
\usepackage[switch]{lineno}
\usepackage{color,soul}

\newcommand\vsini{v\sin{i}}
\newcommand\afe{\rm[\alpha/Fe]}
\newcommand\mh{\rm[M/H]}
\newcommand\logg{\log{g}}
\newcommand\teff{T_{\rm eff}}
\newcommand\kms{\rm km\,s^{-1}}
\newcommand\bell{B_{\ell}}
\def\nstar{6}

\usepackage{xcolor}

\usepackage{placeins}

\begin{document}

   \title{Rotational modulation and long-term evolution of the small-scale magnetic fields of M dwarfs observed with SPIRou}

   \author{P. I. Cristofari
          \inst{1, 2, 3}
          \fnmsep\thanks{cristofari@strw.leidenuniv.nl}
          \and
          J.-F. Donati\inst{3}
          \and
          S. Bellotti\inst{1,3}
          \and
          \'E. Artigau\inst{4}
          \and
          A. Carmona\inst{5}
           \and
           C. Moutou\inst{3}
           \and
		  X.~Delfosse\inst{5}
		  \and
		  P.~Petit\inst{3}
         \and
         B.~Finociety\inst{3,6}
         \and
         J.~Dias~do~Nascimento\inst{2,7}
          }

   \institute{
            Leiden Observatory, Leiden University,
PO Box 9513, 2300 RA Leiden, The Netherlands
\and 
   	Center for Astrophysics | Harvard \& Smithsonian,
60 Garden Street,
Cambridge, MA 02138, United States
         \and
             Univ. de Toulouse, CNRS, IRAP, 14 avenue Belin, 31400 Toulouse, France
        \and
		Institut Trottier de recherche sur les exoplan\`etes, D\'epartement de Physique, Universit\'e de Montr\'eal, Montr\'eal, Qu\'ebec, Canada
        \and
        Univ. Grenoble Alpes, CNRS, IPAG, 38000 Grenoble, France
        \and
        ACRI-ST, 260 route du Pin Montard, BP 234, 06904 Sophia-Antipolis (France)
        \and
        Departamento de F\'isica, UFRN, CP 1641, 59072-970, Natal, RN, Brazil
             }

   \date{Received Month DD, YYYY; accepted Month DD, YYYY}
 
  \abstract
   {M dwarfs are known to host magnetic fields, impacting exoplanet studies and playing a key role in stellar and planetary formation and evolution. Observational constraints are essential to guide theories of dynamo processes believed to be at the origin of those fields, in particular for fully-convective stars whose internal structure differs from those of partially-convective stars. Observations revealed long-term evolution of the large-scale magnetic field reconstructed with Zeeman-Doppler imaging, and a diversity of their topologies. These large-scale magnetic fields, however, only account for a small amount of the unsigned magnetic flux at the stellar surface that can be probed by directly modeling  the Zeeman broadening of spectral lines in unpolarized spectra.
   }
   {We aim at investigating the long-term behavior of the average small-scale magnetic field of fully-convective and partially convective M dwarfs with time, and assess our ability to detect rotational modulation and retrieve rotation periods from time series of field measurements derived from unpolarized spectra.}
   {We perform fits of synthetic spectra computed with \texttt{ZeeTurbo} to near-infrared high-resolution spectra recorded with SPIRou between 2019 and 2024 in the context of the SLS and SPICE large programs. The analysis is performed on  the spectra of 2 partially convective (AD Leo, DS Leo) and 3 fully convective (PM~J18482+0741, CN~Leo, Barnard star) M dwarfs, along with EV~Lac whose mass is close to the fully-convective limit. Our analysis provides measurements of the average {small-scale} magnetic field, which are compared to longitudinal magnetic field and temperature variation measurements (d$Temp$) obtained from the same data.}
   {We were able to detect the rotation period in the small-scale magnetic field series for 4 of the \nstar\ stars in our sample. We find that the average magnetic field can vary by up to 0.3\,kG throughout the year (e.g., CN~Leo), or of up to 1\,kG across rotation phases (e.g., EV~Lac). 
   {The rotation periods retrieved from longitudinal and small-scale magnetic fields are found in agreement within error bars.}
   d$Temp$ measurements are found to anti-correlate with small-scale magnetic field measurements for three stars (EV~Lac, DS~Leo and Barnard's star). 
   }
   {The results demonstrate our ability to measure rotation periods from high-resolution data from small-scale magnetic field measurements, provided that the inclination of the observed targets is sufficiently large.
   We observe long-term fluctuations of the average magnetic field that could indicate magnetic cycles in the {parent dynamo} processes. These {long-term variations} appear mainly uncorrelated with large-scale magnetic field variations probed through the longitudinal field measurements. 
   Large variations in amplitude of the rotationally modulated signals, in particular, hint towards a change in the distribution of the surface inhomogeneities accessible to Zeeman broadening measurements.
   }
     
    \titlerunning{Rotational modulation and long-term evolution of small-scale magnetic fields} \authorrunning{P. I. Cristofari}
   \keywords{Techniques: spectroscopic --
	Stars: low-mass -- 
	Stars: magnetic field
}

   \maketitle
  
\section{Introduction}

M dwarfs have attracted increasing attention in the past years for their properties as planet hosts. The characterization of M dwarfs is, in particular, essential to establish reliable constraints on the detected planets~\citep{bonfils-2013, dressing-2015, gaidos-2016}. 
 M dwarfs are known to host magnetic fields~\citep[e.g.,][]{saar-1985, johns-krull-1996, shulyak-2014, kochukhov-2021, reiners-2022}, routinely impairing exoplanet characterization, and at the origin of activity phenomena leading to spurious signals in radial velocity curves~\citep[e.g.,][]{dumusque-2021, bellotti-2022}. These fields play a crucial role in stellar formation and evolution~\citep{donati-2009}, and are responsible for a number of observable phenomena, such as momentum loss over the stellar life~\citep[e.g.,][]{skumanich-1972, vidotto-2014} that led to the development of gyrochronology~\citep{barnes-2003}, or surface inhomogeneities (spots, plages, faculae). 

M dwarfs are the most numerous stars in the solar vicinity~\citep{reyle-2021}, with masses ranging from 0.08 to 0.57\,M$_\odot$~\citep{pecaut-2013}. 
Stars with masses~${<0.35}$\,M$_\odot$ are predicted to be fully convective.
In partially convective stars, with M>0.35\,M$_\odot$, the tachocline~\citep[region at the interface between the inner radiative core and outer convective envelope,][]{chabrier-1997} has been proposed as a key ingredient in dynamo processes at the origin of magnetic fields. In fully convective stars, dynamo theories can no longer rely on the tachocline, and an alternative $\alpha^2$-dynamo mechanism is proposed to generate strong magnetic fields without the need of the tachocline~\citep{chabrier-2006, yadav-2015}.

In the last decade, the advent of new high-resolution near-infrared instruments, including spectrometers such as CARMENES~\citep{quirrenbach-2014} and spectropolarimeters such as CRIRES+~\citep{dorn-2023} or SPIRou~\citep{donati-2020}, along with the development of new spectral modeling codes, have provided new constraints on the atmospheric properties of a large number of M dwarfs~\citep[e.g.,][]{rajpurohit-2018, passegger-2019, marfil-2021, sarmento-2021, cristofari-2022, cristofari-2022b}. The polarized data recorded with such instruments led to numerous studies focused on the large-scale magnetic field of such stars~\citep[e.g.][]{finociety-2023, donati-2023a, bellotti-2024}, while several other works relied on unpolarized spectra to estimate the average surface magnetic field of low-mass and Sun-like stars by modeling the Zeeman broadening and intensification of well-selected spectral lines~\citep{shulyak-2017, reiners-2022, cristofari-2023, cristofari-2023b, kochukhov-2024}. Those measurements open the door to complementary studies aiming at drawing a complete picture of the magnetic fields in M dwarfs from large-scale and small-scale field measurements~\citep[e.g.][]{kochukhov-2017, donati-2023a}. Large surveys of magnetic M dwarfs have produced timeseries of spectra for hundred of stars, providing the data necessary to investigate the link between rotation and magnetic fields. Magnetic activity is known to scale with Rossby number, defined as the ratio  between rotation period and  convective turnover time, and recent works found that magnetic fields follow a similar trend for fully and partially convective stars~\citep[e.g.][]{reiners-2022, cristofari-2023b}.

In this paper, we present for the first time a systematic investigation of the evolution of the average magnetic field and longitudinal field on periods of months or years for several magnetic M dwarfs observed with SPIRou.
We rely on the long-term monitoring of 6 fully-convective and partially-convective targets (EV~Lac, DS~Leo, CN~Leo, PM~J18482+0741, AD~Leo and Barnard star) with SPIRou.
We introduce the observations in Sec.~\ref{sec:observations} and the tools used for the analysis in Sec~\ref{sec:analysis}. In Sec.~\ref{sec:results-nights} we present our results, before discussing them in Sec.~\ref{sec:discussion}.

\section{Observations and reduction}
\label{sec:observations}

The analysis presented in this paper relies of observations recorded with SPIRou, the spectropolarim\'etre infrarouge installed at the Canada-France-Hawaii Telescope, in the context of the SPIRou legacy Survey (SLS), a three year program that was allocated 310 nights between 2019 and 2022, of the SPIRou Legacy Survey - Consolidation and enhancement (SPICE) large program, and of observations obtained in the context of PI programs (Run ID 24AC25, 23AD98, 22BF10, 24AF17, 24BF13, and 23BF08). We focus on a sample of 5 strongly magnetic M dwarfs (EV Lac, AD Leo, DS Leo, CN Leo, PM J18482+0741, see Table~\ref{tab:literature}), to which we add one quiet star (Barnard's star).
Spectra were reduced with \texttt{APERO} version 0.7.291~\citep{cook-2022}. Wavelength calibration, blaze estimation from flat-field exposures and telluric correction were performed by \texttt{APERO}.

The unpolarized spectral orders were normalized using a low-degree polynomial, corrected for the barycentric Earth radial velocity (BERV), and re-binned on a common wavelength grid using a cubic interpolation. In addition to the spectra recorded each night, we compute a median spectrum, referred to as `template' in the rest of the paper, by taking the median of each pixel in the barycentric reference frame. For those templates, the signal-to-noise ratio (SNR) per 2\,$\kms$ pixel can reach up to 2000 in the $H$ band.

The polarimetric data were also reduced with the \texttt{LIBRE-ESPRIT} package adapted to SPIRou data~\citep{donati-2020}. The polarimetric products were used to compute mean Stokes I and V profiles with least squares deconvolution~\citep[LSD, ][]{donati-1997b}. 
The longitudinal field estimates ($B_\ell$) obtained with these tools were compared to those obtained from the polarimetric products of \texttt{APERO} and the publicly available code LSDpy\footnote{https://github.com/folsomcp/LSDpy}. The two datasets provide similar results, and we therefore rely on the extensively tested \texttt{LIBRE-ESPRIT} implementation which leads to slightly smaller error bars.

\begin{table*}
	\caption{Parameters retrieved in the literature for the stars in our sample.}
	\label{tab:literature}
 \centering
	\begin{tabular}{ccccccccc}
		\hline
		\hline
		Star & Gliese ID & Spectral type & $P_{\rm rot}$  (d) & $M$ ($M_{\sun}$) & $R$ ($R_{\sun}$) & $\tau$ (d) & $R_{\rm O}$ \\ 
		\hline
		DS Leo & Gl 410 & M1.0V & $14.0\pm0.1^{1}$ & $0.57\pm0.02$ & $0.53\pm0.02$ &
              $38 \pm 23$ & $0.37 \pm 0.23$ \\
		AD Leo & Gl 388 & M3V & $2.2399\pm0.0006^{2}$ & $0.42\pm0.02$ & $0.39\pm0.02$ &  
            $57 \pm 33$ & $0.04 \pm 0.02$ \\
		EV Lac & Gl 873 &  M4.0V & $4.3715\pm0.0006^{2}$ & $0.32 \pm0.02$ & $0.31\pm0.02$ & 
                $76 \pm 42$ & $0.06 \pm 0.03$ \\
	      Barnard's star & Gl~699 & M4V & $136\pm13^3$ & $0.16\pm0.02$ & $0.185\pm0.004$ & 
                   $125 \pm 67$ & $1.09 \pm 0.59$ \\
		PM J18482+0741 & ... &  M5.0V &  $2.76\pm0.01^{4}$ & $0.14\pm0.02$ & $0.17\pm0.02$ &  
              $134 \pm 71$ & $0.02 \pm 0.01$ \\
		CN Leo & Gl 406 & M6V & $2.704\pm0.003^{4}$ &  $0.11\pm0.02$  & $0.13\pm0.02$ & 
              $147 \pm 79$ & $0.02 \pm 0.01$ \\
		\hline
	\end{tabular}
 \tablefoot{Masses and radii were taken from~\citet{cristofari-2023, cristofari-2023b}. Convective turnover times ($\tau$)  were computed from the mass with the relation introduced in~\citet{wright-2018}. The Rossby number is defined as $R_{\rm O} = P_{\rm rot}/\tau$. \\
 Ref. -- 
	 (1)~\citet{donati-2008},
		(2)~\citet{morin-2008},
		(3):~\citet{donati-2023b}, 
		(4):~\citet{diez-alonso-2019}, 
	
  }
\end{table*}

\section{Data analysis}
\label{sec:analysis}
\subsection{Stellar spectra modeling}
Our analysis follows the process presented in~\citet{cristofari-2023}, that we briefly describe in this section.

We rely on a grid of synthetic spectra computed with \texttt{ZeeTurbo}~\citep{cristofari-2023} from MARCS model atmospheres~\citep{gustafsson-2008}. Our grid spans temperatures ranging from 2700 to 4000\,K, $\logg$ from 3.5 to 5.5 and $\mh$ from -0.75 to 0.75\,dex. For each set of atmospheric parameters, spectra were computed for surface magnetic fields ranging from 0 to 10\,kG in steps of 2\,kG, assuming that the magnetic field is radial in all points of the photosphere. 
Every spectrum in the grid was computed assuming local-thermodynamic equilibrium (LTE). 
We rely on the results of~\citet[][see Fig.~10 therein]{wende-2009} to determine a microturbulent velocity--$\teff$ relation, yielding $v_{\rm mic} = a\,\teff^2+b\,\teff+c$, with $a=4.4\times10^{-7}\,\kms K^{-1}$, $b=-2.4020\times10^{-3}\,\kms K^{-1}$ and $c=3.5191\,\kms K^{-1}$. The $v_{\rm mic}$ considered in our grid then range from $\sim 0.25$ to $0.98\,\kms$. We note that previous tests showed that imposing $v_{\rm mic}=1\,\kms$ for all spectra within the grid has a very small impact on the results of our current analysis.

We adopt the same convention as in~\citet{cristofari-2023, cristofari-2023b} and model the stellar spectra with a linear combination of the synthetic spectra computed for various magnetic field strengths so that $S=\sum f_iS_i$, with $S_i$ the spectrum for magnetic field $i$\,kG and $f_i$ the filling factor associated to that component. Our fitting procedure ensures that $\sum f_{i}=1$. To obtain the best fit to the data, we rely on Markov Chain Monte Carlo (MCMC) with  the log-likelihood explicited in~\citet{cristofari-2023b}, and derive the parameters from the posterior distributions.

\subsection{Atmospheric characterization}

In a first step, we derive the atmospheric parameters and average magnetic fields obtained for each star by applying our process to the templates (see Table~\ref{tab:stellar_characterization}). This analysis was performed with the same assumptions on $\vsini$ than~\citet{cristofari-2023}, except for Barnard's star, whose $\vsini$ is set to 0~$\kms$~\citep{cristofari-2023b}. The retrieved atmospheric properties are in excellent agreement with those derived in our previous study~\citep{cristofari-2023} performed on earlier version of the templates. Note that we obtain $\logg\approx5.0$\,dex for PM~J18482+0741 and CN~Leo, larger than in~\citet{cristofari-2023}, and closer to the values expected from radius and mass estimates ($5.12\pm0.13$\,dex and $5.25\pm0.17$\,dex, respectively, see Table~\ref{tab:literature}). Those improvements can be attributed to our refined line list selection, and improvements to the normalization functions. Surface gravity is known to be difficult to constrain for M dwarfs, and line selection has a significant impact on the results. The error bars on reported atmospheric parameters were inflated to account for some of the systematics following~\citet{cristofari-2022}.

In a second step, we fix the atmospheric parameters to those derived from the template, and estimate $\langle B \rangle$ from the spectra recorded each night, obtaining a time series of magnetic field measurements for each star.
For all stars but Barnard's star, we model the spectra with a combination of magnetic models ranging from 0 to 10\,kG in steps of 2\,kG. For Barnard's star, only the 0 and 2\,kG components were considered.

\begin{table*}
\center
\caption{Atmospheric parameters and small-scale magnetic field. }
\label{tab:stellar_characterization}
\begin{tabular}{ccccccccc}
\hline
\hline
Target & $\teff$ (K) & $\logg$ (dex) & $\mh$ (dex) & $\vsini$ ($\kms$) & $i$~($^{\circ}$)$^{a}$ & $\zeta_{\rm RT}$ ($\kms$)$^b$ & $\langle B \rangle$ (kG) \\
\hline
DS Leo & {$3797\pm30$} &  {$4.67\pm0.05$} &  {$-0.02\pm0.10$} & $1.5$ & $51\pm48$ & {$3.01\pm0.07$} & {$0.79\pm0.02$}\\ 
AD Leo & {$3475\pm30$} & {$4.81\pm0.05$} & {$0.23\pm0.10$} & $3.0$ & $20\pm11$ & {$2.19\pm0.15$} & {$3.13\pm0.05$}\\ 
EV Lac & {$3342\pm30$} & {$4.75\pm0.05$} & {$0.02\pm0.10$} & $3.0$ &   $57\pm57$ & {$4.22\pm0.14$} & {$4.54\pm0.09$}\\ 
Barnard's star & {$3300\pm31$} & {$4.71\pm0.06$} & {$-0.54\pm0.10$}$^{c}$ & $<0.1^{d}$ &  ... & {$3.79\pm0.17$} & {$0.51\pm0.08$}\\
PM~J18482+0741 & {$3102\pm32$} & {$4.97\pm0.06$} & {$0.01\pm0.10$} & $2.4$ & $40\pm27$ & {$5.51\pm0.18$} & {$1.27\pm0.14$}\\ 
CN Leo &  {$2912\pm31$} & {$5.00\pm0.07$} & {$0.22\pm0.11$} & $2.0$ & $55\pm45$  &  {$4.94\pm0.29$} & {$3.08\pm0.26$}\\ 
\hline
\end{tabular}
 \tablefoot{Projected rotational velocities ($v\sin{i}$) for our analyses were taken from~\citet{morin-2008} for AD~Leo and~\citet{reiners-2018} for PM~J18482+0741. For CN~Leo, EV~Lac and DS~Leo, $v\sin{i}$ estimates were taken from~\citet{cristofari-2023}, who revised some values based on rotation periods and radii. Inclinations were derived from $v\sin{i}$, $P_{\rm rot}$ and radii. \\
 	{
	 \textit{a}:~Inclinations computed assuming a $1.0$\,$\kms$ uncertainty on $v\sin{i}$.\\
	  \textit{b}:~In the present analysis, we fixed $v\sin{i}$ and fit $\zeta_{\rm RT}$. Consequently, broadening that would arise from non-physical or unidentified sources may lead to larger $\zeta_{\rm RT}$ estimates.\\
	 \textit{c}:~For Barnard's star, we additionally fit for the $\afe$ parameter~\citep[see][]{cristofari-2022b}, yielding: $\afe=0.09\pm0.10$\,dex.\\
	 \textit{d}:~Maximum $v\sin{i}$ assuming an inclination of $90^{\circ}$.\\
	}
}
\end{table*}

\section{Temporal modulation of the small-scale and large-scale magnetic fields}
\label{sec:results-nights}

\subsection{Quasi-periodic Gaussian Process fit to the data}

Quasi-periodic Gaussian Processes (GP) have become popular to study the evolution of activity of stars and extract rotation periods~\citep[see, e.g.,][]{fouque-2023, donati-2023a, donati-2023b}.

We investigate the temporal modulation of our data by using a GP regression with a kernel used in previous studies~\citep{angus-2018, fouque-2023}:
\begin{equation}
\kappa(t_i, t_j)=\alpha^2\exp\Big[-\frac{(t_i - t_j)^2}{2l^2}-\frac{1}{2\beta^2}\sin^2\Big(\frac{\pi(t_i - t_j)^2}{P_{\rm rot}}\Big)\Big] + \sigma^2\delta_{ij}
\label{eq:equation}
\end{equation}
With $\alpha$ the amplitude of the GP, $\beta$ a smoothing factor, $l$ the decay time, $\sigma$ the standard deviation of an added uncorrelated white noise and $P_{\rm rot}$ the recurrence period, here assimilated to the rotation period of the star.
This flexible model allows one to model complex structures, but can be degenerate, resulting in over-fitting~\citep{angus-2018}. For each star, we therefore obtain a fit with wide priors (assuming little knowledge of the GP hyper-parameters).

Our implementation\footnote{\url{https://github.com/pcristof/star-activity-tools}} relies the \texttt{George}~\citep{george-gp} Python package and a workflow loosely based on that of~\citet{martioli-2022}. Our tools explore the parameter space with a Markov Chain Monte Carlo (MCMC), relying on the emcee package~\citep[][]{mackey-2008} to estimate the set of hyper-parameters {(see equation~\ref{eq:equation}) and mean value ($\mu$)} that lead to the highest likelihood described in~\citet{rasmussen-2006, foreman-2017} and implemented in \texttt{George}.

For most targets, the minimum reduced $\chi^2$ ($\chi^2_r$) of the GP fit is much lower than one. These likely reflect that the error bars computed for on $\langle B \rangle$ attempt to account for some of the systematics associated with the measurements, and do not reflect relative precision. Given that we are investigating relative variation of the measurements, we re-scale the error bars using the dispersion of the residuals in order to ensure that the minimum reduced $\chi^2_r$ is closer to 1. 
In the rest of the papers, the errors listed for $\langle B \rangle$ correspond to these re-scaled error bars.

To estimate the parameters leading to the best fit to the data, we search for the combination of hyper-parameters maximizing the likelihood. In order to avoid bias by a single walker's position, we consider the top 1\% walkers with highest likelihood, and take the median of these walkers parameters. Our process was applied to $\langle B \rangle$ and $B_{\rm \ell}$ measurements. We additionally performed simultaneous GP fits on the series of $\langle B \rangle$ and $B_\ell$ measurements, with the same rotation period for both GPs.
Last but not least, we applied our process to d\textit{Temp} measurements obtained following~\citet{artigau-2024}, {attempting to capture temperature variations at the stellar surface by relying on changes in stellar line profiles throughout the observation campaign. 
For each star, we collected the d$Temp$ measurements obtained with a reference temperature closest to those listed in Table~\ref{tab:stellar_characterization} (i.e. 4000\,K for DS~Leo, 3500\,K for AD~Leo, EV~Lac, and Barnard's star, and 3000\,K for PM~J18482+0741 and CN~Leo).
The minimum reduced $\chi^2_r$ of the best GP fit to the series d$Temp$ measurements range from 2 to 94, with particularly large values for CN~Leo and PM~J18482+0741 ($\chi^2_r=57$ and $\chi^2_r=94$, respectively), likely indicating that the error bars are under-estimated. For d$Temp$, we therefore re-scaled error bars to ensure that the minimum reduced $\chi^2$ is closer to 1 for the best fits.}
The results of the GP fits are presented in Tables~\ref{tab:gp-results-gl410},~\ref{tab:gp-results-gl388},~\ref{tab:gp-results-gl873},~\ref{tab:gp-results-gl699},~\ref{tab:gp-results-pmj} and~\ref{tab:gp-results-gl406}.

\begin{figure*}
	\centering
	\includegraphics[width=.99\linewidth]{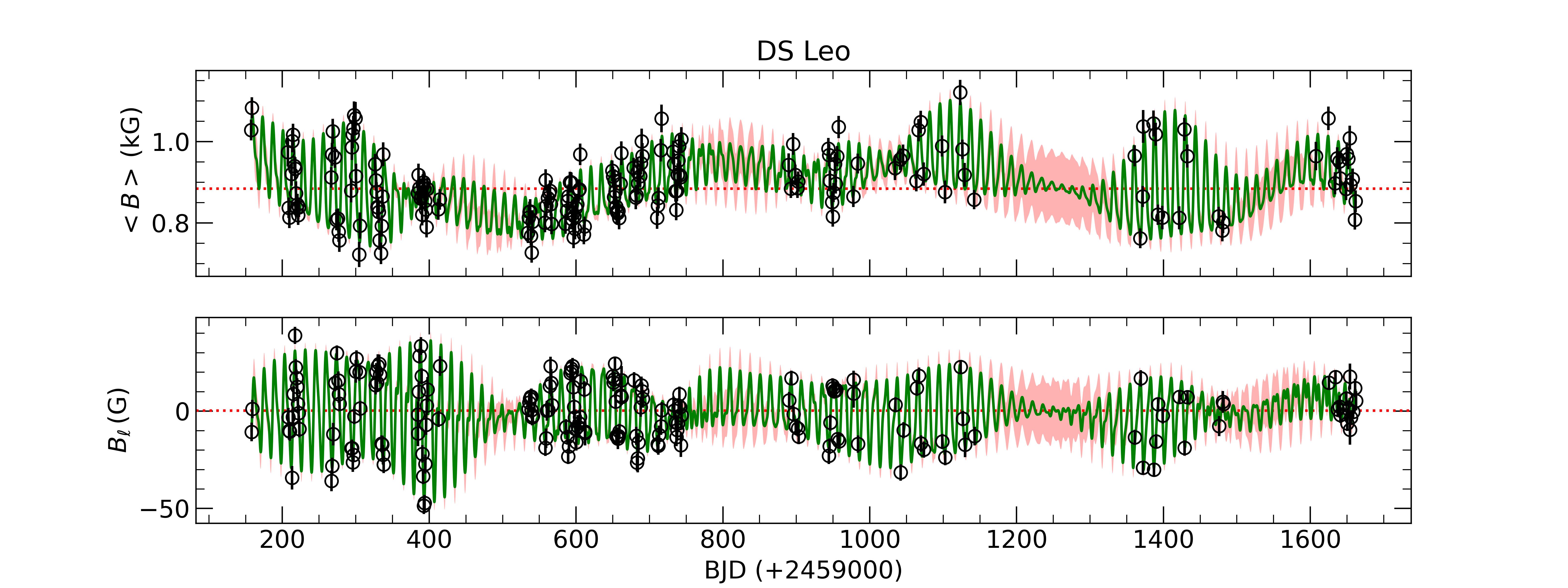}
	\caption{Simultaneous fit of two GPs on our small-scale and large-scale magnetic fields measurements. The rotational velocity $P_{\rm rot}$ is the same for the two GPs. The pink shaded area shows the uncertainty on the GP fit.}
	\label{fig:dsleo-multi-inline}
\end{figure*}

\begin{figure}
\includegraphics[width=.9\linewidth]{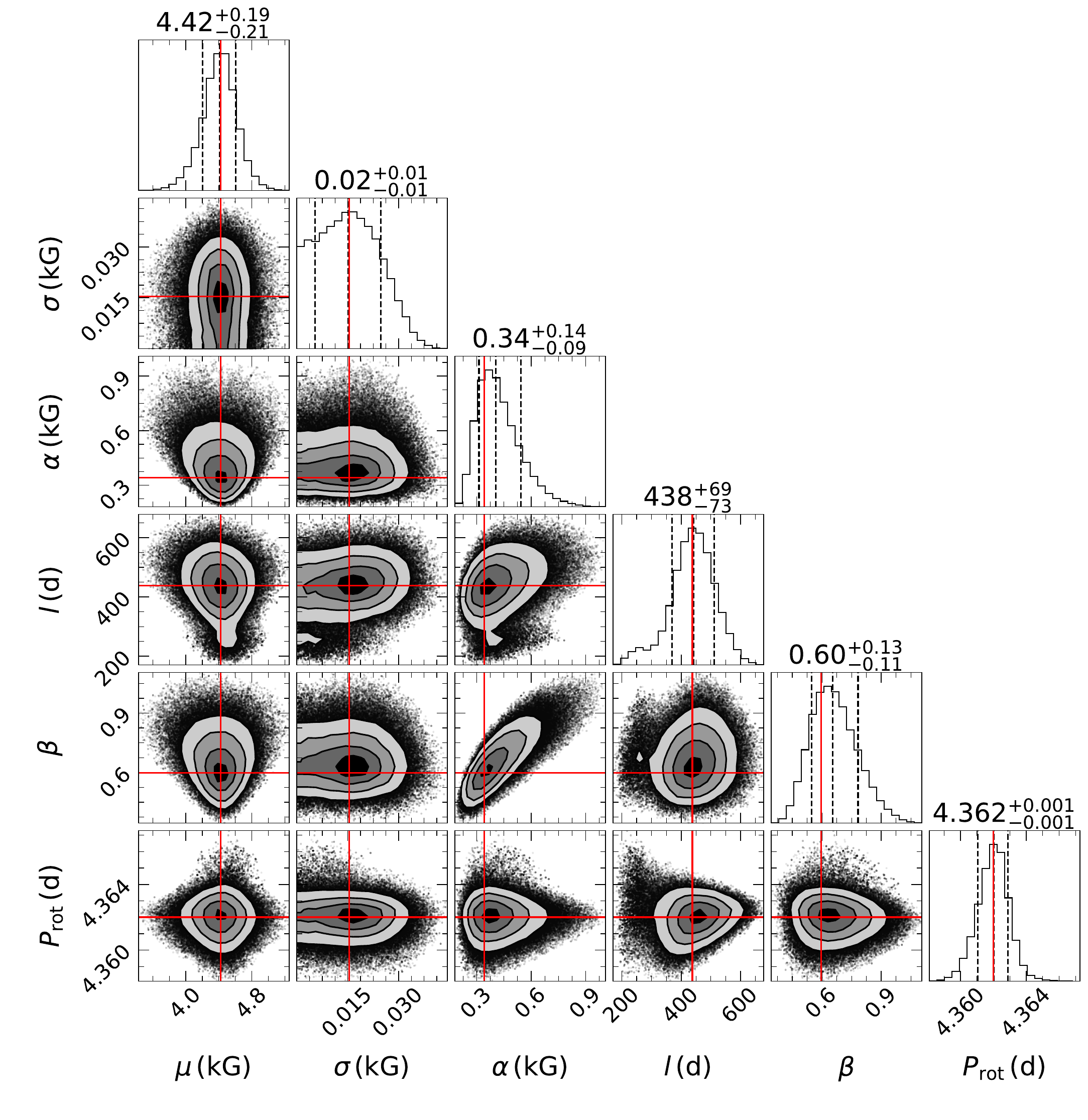}
\caption{Posterior distribution of the hyper-parameters obtained for EV~Lac. Red lines mark the median of the 1\% of walkers which have the highest likelihood.}
\label{fig:corner-evlac}
\end{figure}

\begin{figure*}
\centering
\includegraphics[width=.90\linewidth]{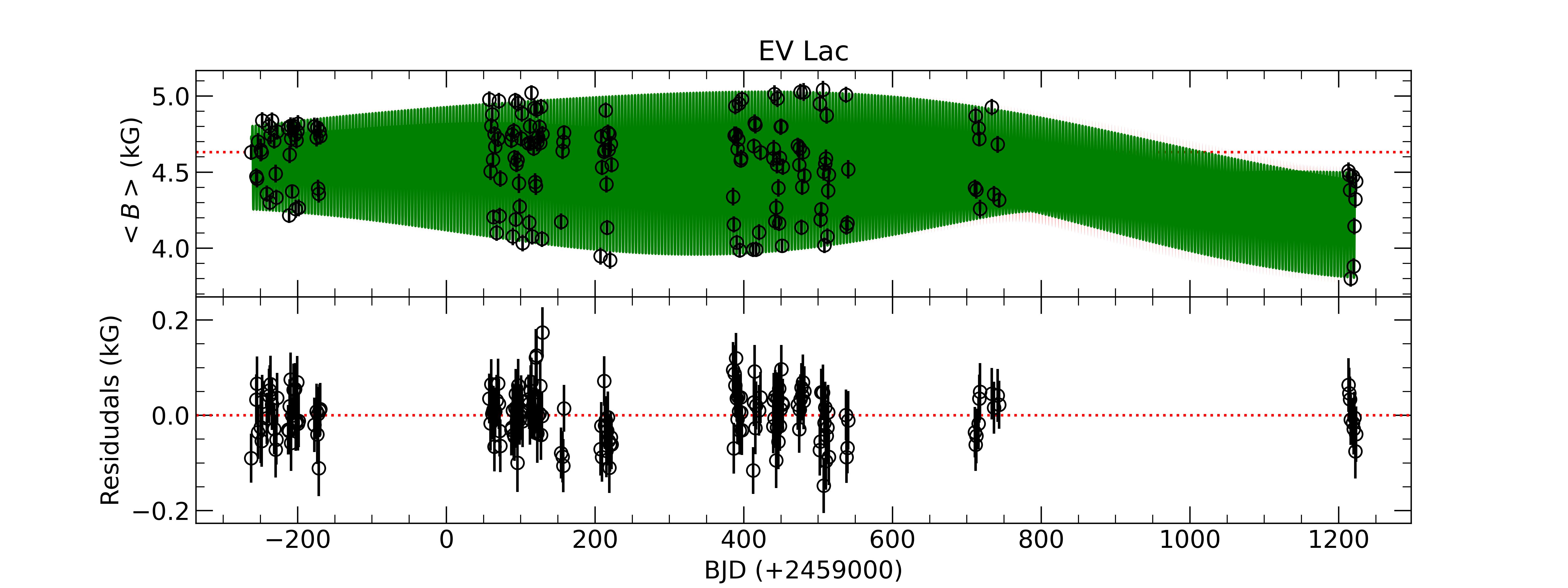}
\includegraphics[width=.90\linewidth]{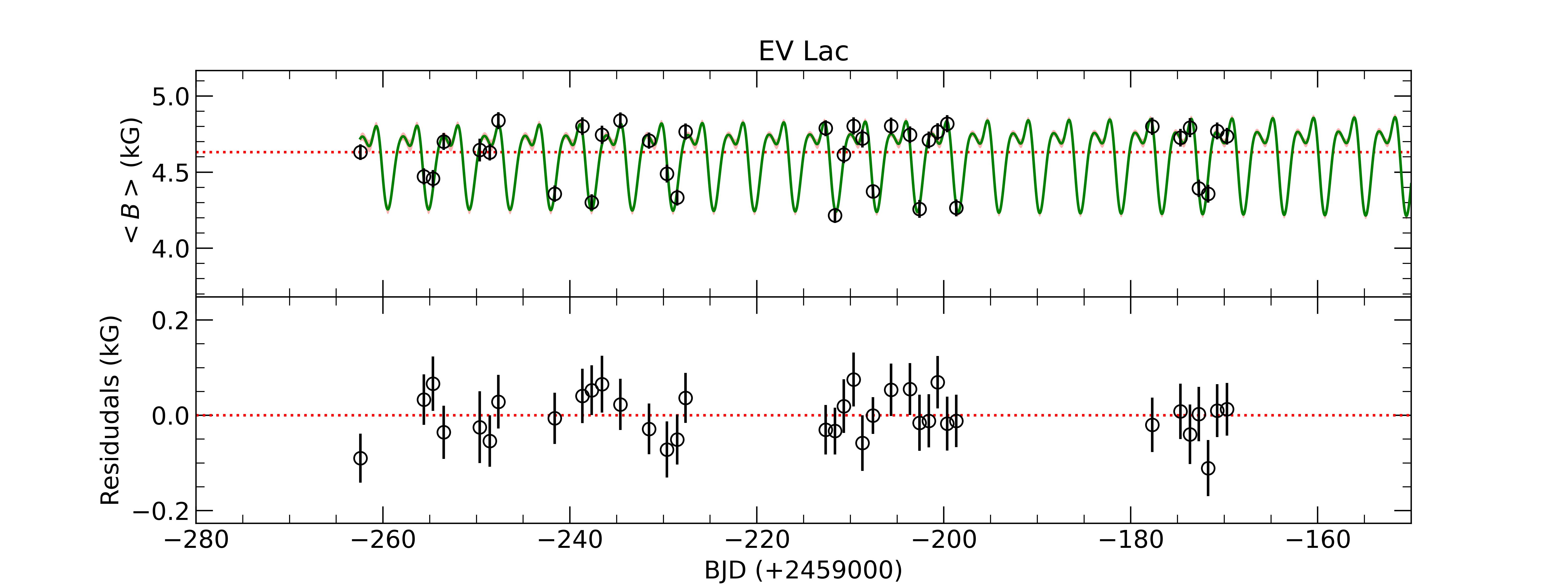}
\includegraphics[width=.90\linewidth]{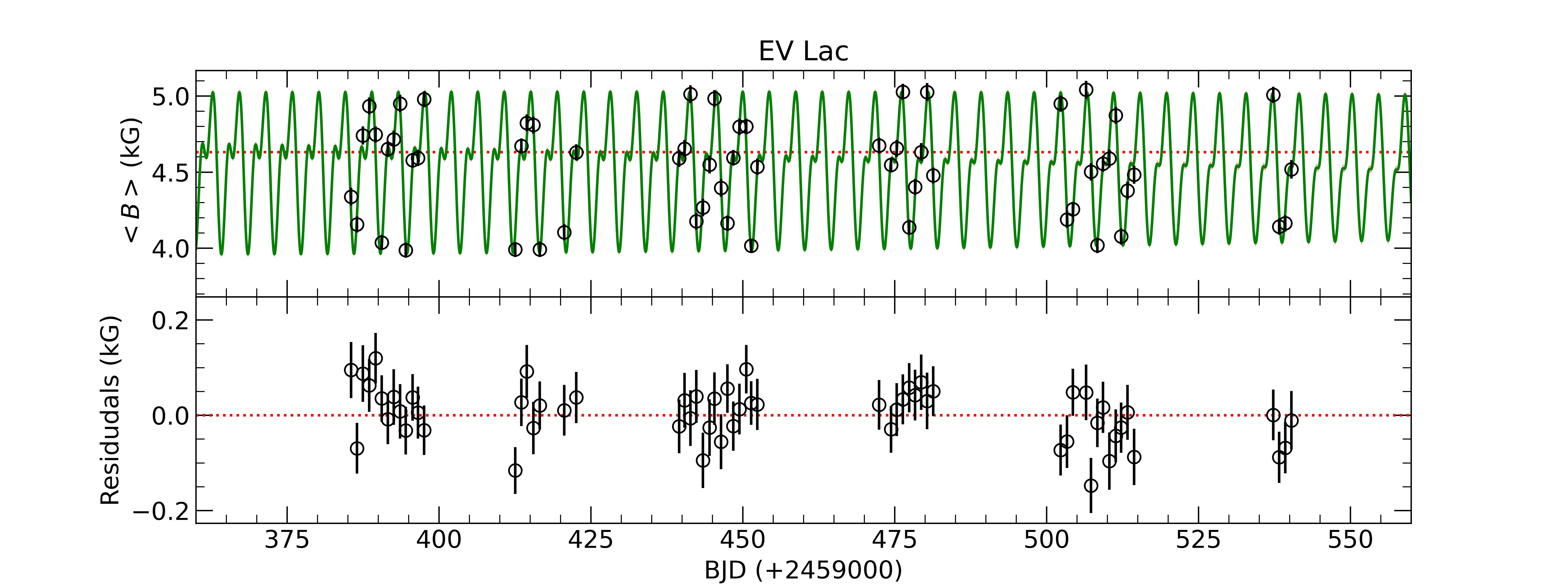}
\caption{Best fit GP fit (green) obtained on our small-scale magnetic fields measurements (black circles). The top panel shows the results obtained over our entire data set, while the middle and bottom panels are zoomed on different observation periods.}
\label{fig:evlac}
\end{figure*}

\subsection{DS Leo (Gl\,410)}

DS~Leo is one of the most massive and least magnetic star in our sample, with an average field of $0.79\pm0.02$\,kG, and around which a possible planetary system was recently detected with these SPIRou data~\citep{carmona-2025}.
We detect a clear modulation of the magnetic field strength, with a detectable signal at $13.953\pm0.088$\,d, consistent with previously reported rotation estimates~\citep{donati-2008}. This period is also in very good agreement with that obtained by applying our process on $B_\ell$ measurements ($14.001\pm0.077$\,d).

We observe a variation in the amplitude of $\langle B \rangle$, with a minimum reached between July 2021 and early-2022 (see Fig.~C.1, available on \href{https://zenodo.org/records/16744873}{Zenodo}). The amplitude of the signal varies by more than 0.3\,kG, going from $\sim$0.9\,kG in March 2021 to $\sim$0.3\,kG in July 2021.
The amplitude in $B_\ell$ measurements is also observed to decrease and then increase, although the minimum is reached in early-2022, and a maximum is reached around July 2021 (see Fig.~D.1, available on \href{https://zenodo.org/records/16744873}{Zenodo}). The series of $\langle B \rangle$ and $|B_\ell|$ appear uncorrelated, yielding a Pearson coefficient of $0.14$.
GP fits obtained on the time series of filling factors show a clear rotational modulation of the 2~kG component as well as for the non-magnetic component. No clear modulation can be found for the other filling factors whose values are close to 0.

DS~Leo was reported to display differential rotation~\citep{hebrard-2016} with rotation periods at the equator and the pole of $13.37\pm0.86$~d and $14.96\pm1.25$~d, respectively. Our period derived from small-scale magnetic field measurements falls between these values.

We carried out a simultaneous fit of two GPs on the series of $\langle B \rangle$ and $B_\ell$ measurements (see Fig.~\ref{fig:dsleo-multi-inline}), yielding a rotation period of $13.980\pm0.059$\,d,  consistent with those derived above. The results of the GP fits are presented in Table~\ref{tab:gp-results-gl410}.

We applied our process to the d$Temp$ measurements obtained for Gl~410 (see Fig.~E.1, available on \href{https://zenodo.org/records/16744873}{Zenodo}). Our best GP fit to the data yields a well constrained recurrence period $P_{\rm rot}=14.214\pm0.106$\,d (see Table~\ref{tab:gp-results-gl410}), slightly larger than that estimated from $\langle B \rangle$ and $B_\ell$.

\subsection{AD~Leo}

Applying our process to our $\langle B \rangle$ measurements for AD~Leo does not provide clear constraints on the rotation period (see Fig.~C.2, available on \href{https://zenodo.org/records/16744873}{Zenodo}).
This result can be attributed to the star's low inclination ($i=19.9\pm1.3^{\circ}$, see Table~\ref{tab:stellar_characterization}): throughout the rotation phase, our processes averages out the signal coming from the pole, with very little modulation in the recorded spectra. Our $\langle B \rangle$ measurements remain relatively stable throughout the months and years and little variation in amplitude or mean is observed. Only the first few and last two data points of our series could hint towards a long term decrease of $\langle B \rangle$.
The longitudinal field of AD~Leo has been shown to vary~\citep{lavail-2018, bellotti-2023}, with a significant increase of the rotationally modulated signal amplitude from mid-2019 to early-2020, and an increase of its mean from early-2020 to mid-2020 (see Fig.~D.2, available on \href{https://zenodo.org/records/16744873}{Zenodo}).

Here again, we perform a simultaneous fit of two GPs to our $\langle B \rangle$ and $B_\ell$ data. This process yields a rotation period of $2.230\pm0.001$\,d, in perfect agreement with that retrieved from $B_\ell$ ($2.230\pm0.001$\,d), suggesting that the constraint on $P_{\rm rot}$ primarily arises from $B_\ell$. The results of the GP fits are presented in Table~\ref{tab:gp-results-gl388}.

We perform a fit of a GP on the d$Temp$ measurements. Relying on wide uniform priors yields $P_{\rm rot}=1.81\pm0.10$\,d. We note that two peaks are visible in the posterior distribution with a secondary peak around the expected rotation period.  To help convergence, we repeated our process fixing the smoothing parameter to $1.5$, and setting a Gaussian prior centered on 2.3 with a 0.3 standard deviation on $P_{\rm rot}$ (see Table~\ref{tab:gp-results-gl388}). With these additional constraints, we retrieve $P_{\rm rot}=2.227\pm0.018$\,d  (see Fig.~E.2 and~G.14, available on \href{https://zenodo.org/records/16744873}{Zenodo}). We note that a $\sim1.7$\,d period is also favored for our GP fit to $\langle B \rangle$ when relying on wide uniform priors, or when setting the decay time to $\beta=0.4$.

\subsection{EV Lac (Gl\,873)}

We carried out our analysis on data recorded for EV~Lac,
whose mass places it very close to the generally adopted fully convective boundary (at about 0.35\,M$\odot$). Recent investigations have reported the presence of two spots at the stellar surface leading to spurious signals in radial velocity measurements (Larue et al. in prep), and a full stokes spectropolarimetric investigation was recently carried out from the SPIRou data secured in September/October 2023~\citep{donati-2025}.
 For this star, the atmospheric characterization yields atmospheric parameters in excellent agreement with~\citet{cristofari-2023}. 
Our process provides well constrained GP hyper-parameters (see Fig.~\ref{fig:corner-evlac}), and a rotation period of $4.362\pm0.001$\,d, very close to that reported by~\citet[][$4.3715 \pm 0.0006$\, d]{morin-2008} and~\citet[][$4.36\pm0.01$\,d]{bellotti-2024}.
This value is also in good agreement with the period retrieved by fitting the GP model to the longitudinal magnetic field measurements ($B_\ell$) obtained from polarimetric data, which yields a rotation period of $4.371\pm0.003$\,d. 

The observations for EV Lac spans $\sim$3 years, and a clear fluctuation of GP amplitude is observed, with a minimum in 2019, and a maximum in 2021 (see Fig.~\ref{fig:evlac}). 
The amplitude of the GP decreases again in 2022, while the observations secured in September/October 2023 yield lower mean small-scale field ($\sim4.1$\,kG against $\sim4.5$\,kG).
The amplitude of $\bell$ and $\langle B \rangle$ follow similar trends, with an increase in amplitude of the $\bell$ for the second and fourth seasons, that appear to decrease again in 2022. Comparing the series of $\langle B \rangle$ and $|B_\ell|$, we compute a Pearson correlation coefficient of $-0.41$.

\begin{figure}
\includegraphics[width=1.0\linewidth]{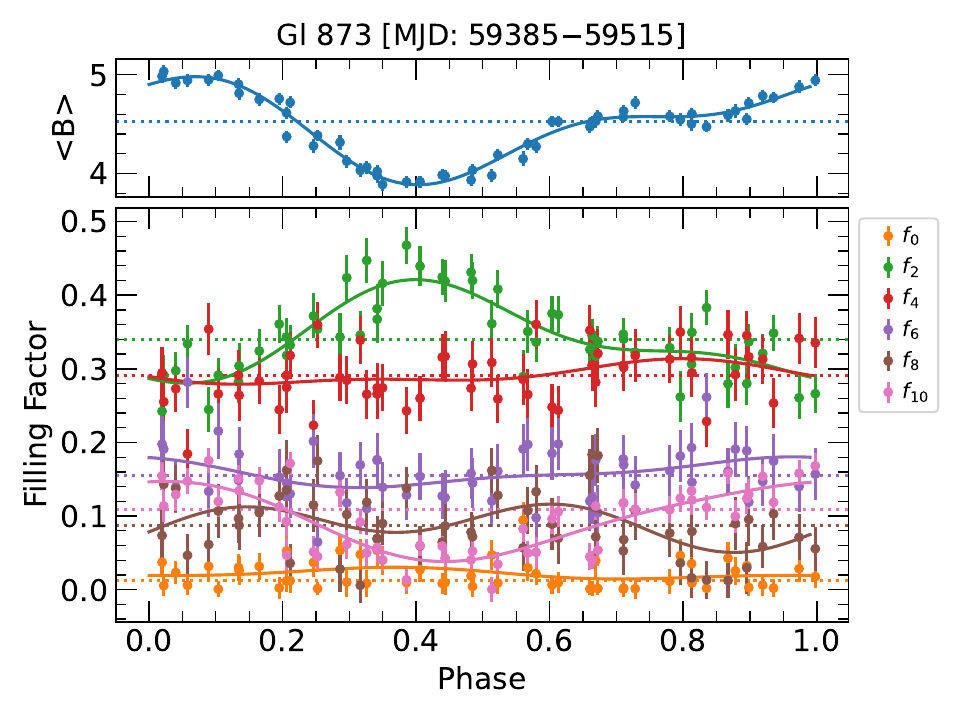}
\caption{Evolution of the filling factors distribution with respect to phase for EV~Lac for observations recorded between MJD 59385 and 59515.
		Solid lines show a fit of a sinusoidal with a first harmonic to the data, assuming the rotation period obtained from our GP fit ($P_{\rm rot}=4.362$\,d, see Sec.~\ref{sec:results-nights}). For each filing factor, the dashed line marks the median of the series of points. Similar figures obtained for DS~Leo, AD~Leo, EV~Lac, Barnard's star, PMJ~J18482+0741 and CN~Leo are presented in Fig.~\ref{fig:b-distrib-gl410},~\ref{fig:b-distrib-gl388},~\ref{fig:b-distrib-gl699},~\ref{fig:b-distrib-pmj}, and~\ref{fig:b-distrib-gl406}.}
\label{fig:b-distrib-gl873}
\end{figure}

The daily sampling of the measurements allows us to observe significant variations in the distribution of the filling factors at different rotation phases (see Fig.~\ref{fig:b-distrib-gl873}). These distributions show a clear increase in the 8~kG component in June 22 and June 26/27 2021 corresponding to days where the magnetic field is close to the average value obtained from the template, while the 10~kG component increases on June 23, 24 or 28 when $\langle B \rangle$ is larger than the value estimated from the template.
To assess this behavior, we fit a GP on individual filling factors rather than $\langle B \rangle$, restricting the priors around the expected rotation period. For EV~Lac, a clear rotation period at {$4.361\pm0.001$\,d and  $4.364\pm0.003$\,d} is detected for the 2 and 10~kG components, respectively. For the other components, convergence of the GP is more challenging and multiple maxima appear in the rotation period histogram.

We repeat the analysis, simultaneously fitting two GPs to the $\langle B \rangle$ and $B_\ell$ measurements, with a common rotation period (see Table~\ref{tab:gp-results-gl873}). The resulting rotation period ($4.363\pm0.001$\,d) remains closer to that estimated from $\langle B \rangle$ alone than that obtained from $B_\ell$. We note that our process applied to $\langle B \rangle$ and $B_\ell$ converges  to significantly different decay times ($421\pm 80$\,d and $125\pm16$\,d, respectively).

We additionally fit a GP to the d$Temp$ measurements for EV~Lac. Our best fit yields $P_{\rm rot}=4.361\pm0.001$\,d, in excellent agreement with that derived from $\langle B \rangle$.   We find that the d$Temp$ obtained from the observations secured in September/October 2023 are on average larger than the measurements obtained on the rest of the data ($\sim4$\,K against $-1$\,K, see Fig.~E.3, available on \href{https://zenodo.org/records/16744873}{Zenodo}), where the small-scale magnetic field becomes weaker.
	We note the strong anti-correlation between $\langle B \rangle$ and d$Temp$ measurements (see Sec.~\ref{sec:dtemp}).

\subsection{Barnard's star (Gl~699)}
To our sample of active M dwarfs, we add Barnard star, a well-known quiet bright M dwarf, with an average magnetic field was estimated to $0.51\pm0.08$~kG (see Table~\ref{tab:stellar_characterization}). Barnard's star has been extensively studied and is relatively bright, leading to high-quality spectra throughout the monitoring campaign. 
This star therefore represents an ideal test case to assess the performance of our method on a slow rotator.

\begin{figure*}
	\centering
	\includegraphics[width=.90\linewidth]{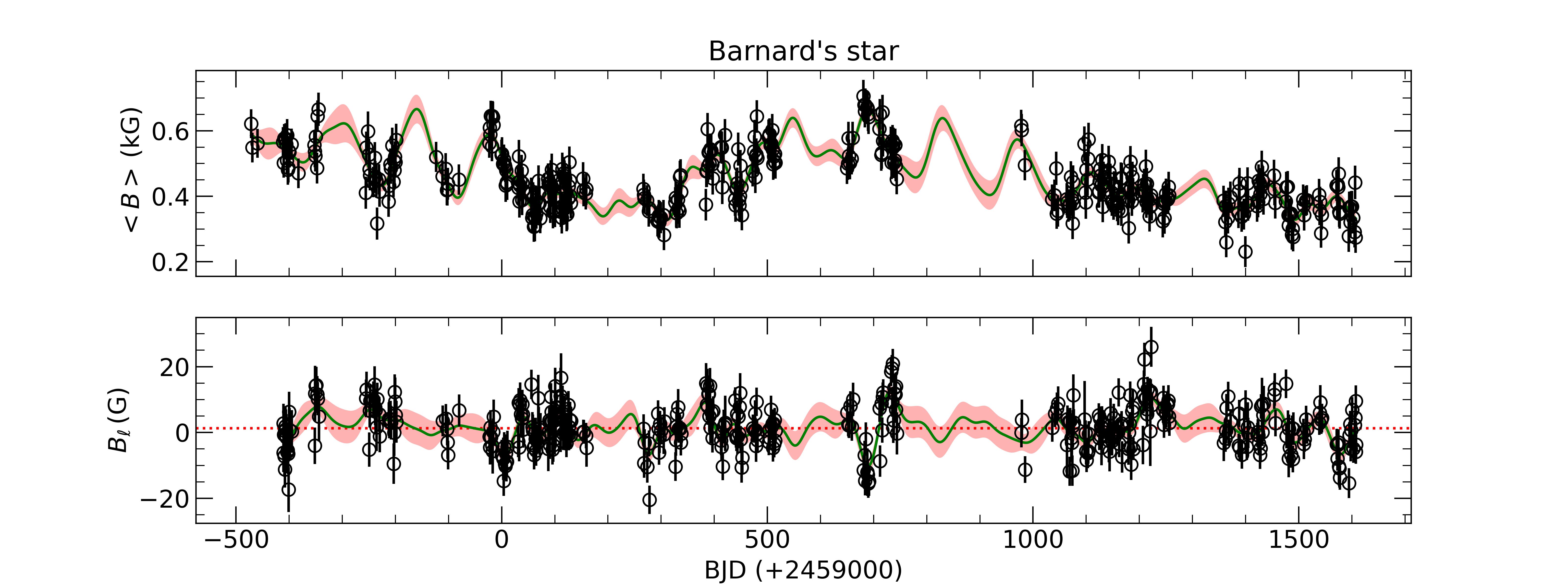}
	\caption{Same as Fig.~\ref{fig:dsleo-multi-inline} for Barnard's star.}
	\label{fig:barnard-multi}
\end{figure*}

From the best obtained GP fit, we retrieved a rotation period of $144\pm6$~d, consistent with the rotation estimated from $B_\ell$ modulation~\citep[$136\pm13$\,d, ][]{donati-2023b} and that reported from d$Temp$ measurements~\citep[$P_{\rm rot}=153\pm3$\,d][]{artigau-2024}.
Those results demonstrate our ability to estimate rotational velocities from small-scale fields of low-activity stars, in spite of the relatively high uncertainty on magnetic field measurements and  long rotation periods.
Relying on the full set of data available for Barnard's star, we note that the convergence of the GP model on $B_\ell$ is challenging. In particular, we find that with wide uniform priors, the GP converges towards a low decay time ($15\pm13$\,d), a $P_{\rm rot}=134\pm50$\,d, and a smoothing factor of $3.0\pm1.9$. Setting the priors so that the decay time be larger than 50 (i.e. $\sim P_{\rm rot}/3$), however, leads to convergence towards a decay time of  $93\pm40$\,d, a smoothing factor of $0.24\pm0.15$ and $P_{\rm rot}=189\pm54$\,d. 
To help convergence, we fixed the decay time to $100$\,d, and the smoothing factor to $0.4$ yielding $P_{\rm rot}=160\pm20$\,d (see Table~\ref{tab:gp-results-gl699}).
Simultaneously fitting two GPs with common $P_{\rm rot}$ to the $\langle B \rangle$ and $B_\ell$  data sets  (see Fig.~\ref{fig:barnard-multi}) yields a rotation period of $139\pm5$\,d.
The series of $\langle B \rangle$ and $|B_\ell|$ measurements appear uncorrelated, with a Pearson correlation coefficient of 0.12.

Here again, we run our process on the d$Temp$ measurements secured for Barnard's star, and obtained a well constrained rotation period $P_{\rm rot}=141\pm12$\,d in excellent agreement with those derived from $\langle B \rangle$ and $B_\ell$  (see Fig.~E.4, available on \href{https://zenodo.org/records/16744873}{Zenodo}).

\subsection{PM~J18482+0741}

PM~J18482+0741 is the second coolest target in our sample and a fully convective star, with a measured average magnetic field of $1.27\pm0.14$\,kG. We first apply our process with uninformed priors on the rotation period. For this star, the constraining $P_{\rm rot}$ is more challenging and multiple peaks are visible in the posterior distribution. To help convergence, we fix the decay time and smoothing to typical values of 300 and 1.50, respectively. We also adopt a Gaussian prior centered on the expected rotation period for this star with a standard deviation of 0.6\,d (see Table~\ref{tab:gp-results-pmj}).
With this prior, our process yields a rotation period of $2.762\pm0.009$\,d. This estimate is in agreement with that derived from $B_\ell$ ($2.761\pm0.001$\,d) and by~\citet[$2.76\pm0.01$\,d]{diez-alonso-2019}.

The mean value of $\langle B \rangle$ varies throughout the monitoring, with a maximum in mid-2022 ($\sim$$1.5$\,kG) and a minimum in mid-2021 ($\sim$$1.2$\,kG). This 0.3\,kG is not negligible compared to the typical error on individual measurements of 0.2\,kG. Furthermore, we observe that the amplitude of the rotationally modulated signal decreases in 2022 compared to the other semesters. For PM~J18482+0741, the $B_\ell$ measurements do not show such long-term variations in amplitude nor mean, and appear uncorrelated to $\langle B \rangle$, with a Pearson coefficient of $0.14$.
The rotational modulation does not clearly appear in when fitting GP to time series of the magnetic filling factors.

Our simultaneous fit of two Gaussian processes to the  $\langle B \rangle$ and $B_\ell$ data yields a rotation period of $2.759\pm0.001$\,d, in very good agreement with that derived from $\langle B \rangle$ and $B_\ell$ (see Table~\ref{tab:gp-results-pmj}).

We carried out our process on the d$Temp$ measurements obtained for PM~J18482+074  (see Fig.~E.5, available on \href{https://zenodo.org/records/16744873}{Zenodo}). The convergence relying on wide uniform priors proves challenging {and multiple peaks are observed in the posterior distribution, impacting our ability to clearly constrain the rotation period. To help convergence, we therefore fixed the smoothing factor to 1.50. A peak close to the highest likelihood is observed at about 2.7\,d, consistent with the $P_{\rm rot}$ derived from $\langle B \rangle$ and $B_\ell$, while several other peaks are observed in the posterior distribution (See Fig.~G.17, available on \href{https://zenodo.org/records/16744873}{Zenodo}).}

\subsection{CN Leo (Gl\,406)}

CN Leo is the coolest star in our sample, and one for which the uncertainties on $\langle B \rangle$ are the largest. These larger uncertainties can be attributed to the large molecular bands impacting the fits to the observed spectra. Using uninformed priors on stellar rotation leads to poor convergence of the GP fit.
To help the fitting process, we fix the decay time and smoothing factor, and adopt a Gaussian prior centered on the expected rotation period (see Table~\ref{tab:gp-results-gl406}). Those additional constraints are not sufficient, however, to clearly detect the rotation period.

We find that our magnetic field estimates drop by $\sim$0.3\,kG from 2019 to 2022 (see Fig~C.6, available on \href{https://zenodo.org/records/16744873}{Zenodo}). Those results could indicate the presence of a long-term periodic fluctuation with a half-period of $\sim$2--3\,yr. 
For this star, a clear rotational modulation is detected in the $B_\ell$ measurements, but the average $B_\ell$ values fluctuates by less than 0.2~kG and does not follow the variations observed in $\langle B \rangle$.
{For CN~Leo, $\langle B \rangle$ and $|B_\ell|$ appear rather uncorrelated, with a Pearson correlation coefficient of $-0.33$.}

We repeat the analysis, simultaneously fitting two GPs to our $\langle B \rangle$ and $B_\ell$ datasets. We find that the process then quickly converges, yielding a rotation period of $2.700\pm0.004$\,d, in excellent agreement with that obtained from $B_\ell$ ($2.696\pm0.006$\,d).

In addition, we perform a GP fit on the d$Temp$ measurements obtained for CN~Leo, fixing the smoothing factor to 1.0 and setting a Gaussian prior on the rotation period to help convergence (see Table~\ref{tab:gp-results-gl406} and Fig.~E.6, available on \href{https://zenodo.org/records/16744873}{Zenodo}). A clear peak at the $P_{\rm rot}=3.003$\,d is observed in the posterior distribution, while a secondary peak at about $2.4$\,d is also observed, de facto impacting the lower uncertainties on this value. We note that these values are not in agreement with that derived from $B_\ell$ measurements.

\section{Comparison with d\textit{Temp}}
\label{sec:dtemp}

\begin{figure}[h!]
	\includegraphics[width=.9\linewidth]{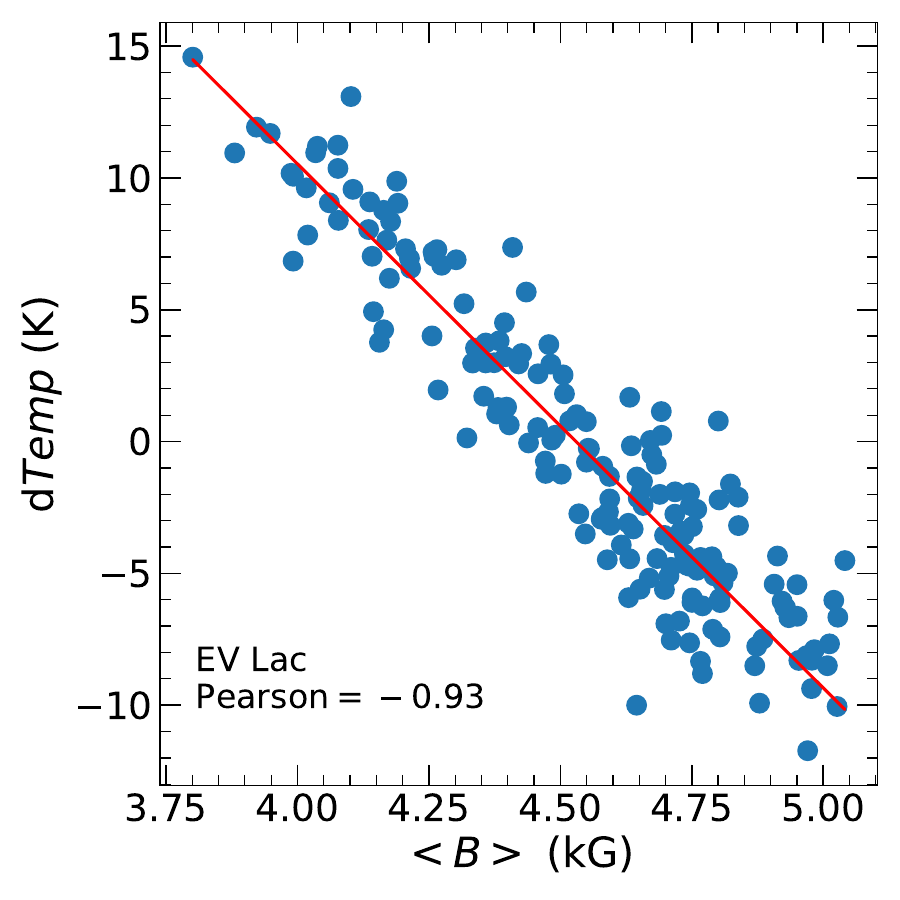}
	\caption{Correlation between $\langle B \rangle$ and d$Temp$ for EV~Lac. {The red line shows a linear with slope $-19.8\pm0.5$\,$\rm K\,kG^{-1}$ and intercept $90.0\pm2.3$\,K.}}
	\label{fig:correlation-gl873}
\end{figure}

We compare our $\langle B \rangle$ estimates to d\textit{Temp} measurements obtained following~\citet{artigau-2024}. 
We find a clear anti-correlation between $\langle B \rangle$ and d\textit{Temp} for DS Leo, EV~Lac, and Barnard's star with Pearson correlation coefficients of $-0.93$, $-0.86$,  and $-0.71$, respectively (see Figures~\ref{fig:correlation-gl873},~H.2, and~H.3).
		 No clear correlation is observed for AD~Leo, CN~Leo and PM~J18482+0741 (correlation coefficients of $-0.45$, $0.26$, and $-0.33$, respectively). For CN~Leo and PM~J18482+0741, no clear rotational modulation is detected in the d$Temp$ measurements, which could result from the lower temperature and stronger molecular bands in their spectra. We also note that~\citet{artigau-2024} obtained a $-0.92$ correlation between $\langle B \rangle$ and d$Temp$ for AU Mic, a M1V star exhibiting strong magnetic behavior.
	
	We note that CN~Leo and AD~Leo, while the convergence of our GP fit to d$Temp$ is more challenging, a clear peak is observed at the expected rotation period in the posterior distribution.
We perform linear fits to the d$Temp$--$\langle B \rangle$ relations for Barnard's star, EV Lac and DS Leo, relying on orthogonal distance regression\footnote{implemented in Scipy}, yielding slopes of $-18.8\pm 0.8$, $-19.8\pm0.5$ and $-60.4\pm2.5$\,$\rm K\,kG^{-1}$, respectively. The intercept of these fits are of $8.2\pm0.3$, $90.0\pm2.3$ and $53.0\pm2.2$\,K for Barnard's star, EV Lac and DS Leo, respectively. These intercepts can be interpreted as the excess in effective temperature one would observe if the stars were not spotted.
	
	We sum the effective temperatures reported in Table~\ref{tab:stellar_characterization} and the intercepts derived above in order to compute the effective temperature of the stars if they were not spotted ($T_{\rm max}$). Relying on the relation by~\citet{berdyugina-2005}, 
	we obtain contrast temperatures (between the photosphere and spots) of about 400, 520,  and 720\,K for Barnard's star, EV Lac and DS Leo, respectively. 
	From $T_{\rm max}$ and the temperature of the spot ($T_{\rm spot}$), we derive a fraction of spot coverage ($f$) for these three stars, assuming that ${T_{\star}=(1-f)\,T_{\rm max}+f\,T_{\rm spot}}$ with $T_{\star}$ the temperature reported in Table~\ref{tab:stellar_characterization}. Our results yield spot coverage fractions of about 2\% for Barnard's star, 19\% for EV Lac, and 8\% for DS~Leo. These results are consistent with EV Lac being the most magnetic star and Barnard's star being the least magnetic star of the three, although the reported fractions should be considered with caution given the numerous sources of uncertainties~\citep{herbst-2021}.

\section{Discussion and conclusions}
\label{sec:discussion}

In this paper, we presented the results of the first investigation of the small-scale magnetic field modulation of several stars. Our process relies on spectral models computed with ZeeTurbo~\citep{cristofari-2023} and high-resolution near-infrared spectra recorded with SPIRou. {Our sample is composed of 3 fully convective stars (CN~Leo, PM~J18482+0741 and Barnard's star), of 2 partially-convective star (AD~Leo, DS~Leo), and of EV~Lac whose mass is close to the fully-convective boundary.}

Our small scale-fields measurements reveal clear rotational modulation for EV Lac, DS Leo, and Barnard's star, with little prior assumptions on rotation period. With more drastic priors, we were able to obtain a rotation period for  PM J18482+0741 but were unable to unambiguously constrain the rotation period of CN~Leo and AD~Leo.
Those results consolidate those previously obtained, for example on AU~Mic~\citep{donati-2023a}, showing that rotational period can be well constrained from $\langle B \rangle$, derived from unpolarized spectra, provided that the inclination  of the targets is sufficiently large and the signal-to-noise ratio of the observations is high. Our sample included the low-activity Barnard's star, for which we retrieved a clear rotation period. These results demonstrate our ability to apply this technique to numerous additional targets monitored over the course of several years, including slow rotators.

In the case of EV~Lac, we find a clear rotational modulation in the 2 and 10\,kG components, while for the other stars the total magnetic flux variations are distributed on more components or restricted to the 2\,kG component. These peculiar results for EV~Lac could arise from the presence of large spots at the stellar surface~\citep[][Larue et al. in prep]{ikuta-2023}, leading to clear modulation of the higher magnetic field component. Subsequent studies will explore the impact of large spots and magnetic topology on the distribution of filling factors.

{We carried out joint fit of two Gaussian Processes to the $\langle B \rangle$ and $B_\ell$ measurements, with a single rotation period, decay time, and smoothing parameter. This approach has the potential to reduce error bars on $P_{\rm rot}$ estimates as the process can rely on additional information. This approach leads to quick and unambiguous convergence for most stars. We note that for PM J18482+0741,  CN~Leo and AD~Leo, the $P_{\rm rot}$ estimates derived with this approach are very close to those obtained from the $B_\ell$ data set alone, with similar or larger error bars. This likely arises from the limited additional constraint provided by the $\langle B \rangle$ data set, for which sub-optimal fits are obtained. For the other stars, however, using both data sets provide additional constraints on the modeling of small-scale and large-scale magnetic fields.}

Our measurements also reveal long-term variations of the small-scale magnetic fields, with amplitudes of the rotationally modulated signal that can increase by a factor of two (e.g. EV~Lac, DS~Leo) or average field strengths that can decrease by up to 0.3\,kG (CN~Leo). {These fluctuations can be larger than the uncertainties computed on $\langle B \rangle$, and are}
therefore likely to impact precision measurements carried out on spectra recorded at a given epoch, or averaged over long periods of time.

We find no clear correlation between the time series of measurements of $\langle B \rangle$ and $|B_\ell|$ for any of the target in our sample. For DS~Leo and EV~Lac, variations in the signal amplitudes appear both in the $\langle B \rangle$ and $B_\ell$ measurements. Only for EV~Lac and AD~Leo do the signals appear moderately anti-correlated or correlated, with Pearson correlation coefficient of $-0.41$ and $0.39$, respectively.
For DS~Leo, although $\langle B \rangle$ and $|B_\ell|$ appear uncorrelated, we note that the amplitude of both rotationally modulated time series vary. These variations appear asynchronous, which could indicate a different relation between the large-scale and small-scale magnetic fields for this star than for EV~Lac.

We compared our obtained small-scale magnetic field measurements to d$Temp$ measurements obtained from the same SPIRou spectra.
No clear correlation between the two sets of measurements was observed for AD~Leo, which is seen pole on, and for Gl~406 and PM~J18482+0741, for which no clear rotational modulation is found in the time series of d$Temp$ measurements.
The comparison revealed clear anti-correlations between the two series of measurements for EV~Lac, DS~Leo and Barnard's star, with Pearson correlation coefficients below $-0.8$.
 	 The anti-correlation between $\langle B \rangle$ and d$Temp$ likely indicates that the presence of strong magnetic fields can give rise to dark spots at the stellar surface. Comparing $\langle B \rangle$ and d$Temp$ for the three stars for which clear anti-correlations are observed, we note that d$Temp$ values vary more quickly with $\langle B \rangle$ for the M1 star (DS~Leo), than for the later type M dwarfs. This observation is further supported by the rapid variation of d$Temp$  with $\langle B \rangle$ reported by~\citet{artigau-2024} for the M1 star AU~Mic. This phenomenon could be a consequence of the variation in spot coverage whose impact is more significant in earlier M dwarfs due to contrast.
 	 For EV~Lac, we observe a significant decrease of the average small-scale magnetic field in September/October 2023. This change is accompanied by an increase in the average d$Temp$ measurements, suggesting that the lower field intensity drives a decrease in cool magnetic regions at the stellar surface. Continued monitoring of EV~Lac will allow us to establish whether these changes are indicative of magnetic cycles and if the small-scale magnetic field has reached a minimum.

	 From linear fits to the  \text{$\langle B \rangle$--d$Temp$} trends, we derived decreasing slopes with increasing contrast, consistent with the previous results suggesting that the contrast between the photosphere and spots increases with temperature~\citep{berdyugina-2005}, and suggesting that our measured small-scale magnetic field arise from spots at the stellar surface. We further deduced the effective temperature of the stars assuming that the obtained $\langle B \rangle$ are caused solely by spots. We relied on the relation proposed by~\citet{berdyugina-2005} to derive spot coverage estimates for Barnard's star, EV~Lac and DS~Leo. These results qualitatively agree with expectations, with the largest spot coverage (19\%) for EV~Lac, and the smallest (2\%) for Barnard's star. The amplitude of the d$Temp$ signals suggest temporal fluctuations of the spot coverage by up to 5\%.
An in depth investigation of the relation between $\langle B \rangle$ and d$Temp$ should attempt to {unambiguously} establish the origin of these anti-correlations.

The results presented in this paper illustrate {how much one can benefit from} the small-scale magnetic fields measurements {derived from nIR spectra collected over} long observation campaigns. Future studies will provide additional measurements of small-scale magnetic fields and their temporal modulation. These measurements, along with longitudinal field measurements and large-scale magnetic field topology reconstructions, will help the community to build a more complete picture of M dwarfs magnetism above and below the fully convective limit. Small-scale magnetic field measurements will also be assets to mitigate their impact on radial-velocity curves~\citep[e.g.,][]{haywood-2016, haywood-2022}, in particular for slow rotating, relatively inactive stars, which are favored targets for the search for habitable exoplanets.

\section{Data availability}
Appendices C, D, E, F and G are available on \href{https://zenodo.org/records/16744873}{Zenodo (https://zenodo.org/records/16744873)}.

\begin{acknowledgements}

This project received funding from the European Research Council (ERC, grant 740651, NewWorlds) under the innovation research and innovation program H2020.

SB acknowledges funding by the Dutch Research Council (NWO) under the project "Exo-space weather and contemporaneous signatures of star-planet interactions" (with project number OCENW.M.22.215 of the research programme "Open Competition Domain Science- M").

XD acknowledges funding from the French National Research Agency in the framework of the Investissements d’Avenir program (ANR-15-IDEX-02), through the funding of the “Origin of Life" project of the Grenoble-Alpes University.

\end{acknowledgements}

  \bibliographystyle{aa} 
  \bibliography{aa54902-25} 

\begin{appendix}

\section{Data tables}

The results of the GP fits are presented in Tables~\ref{tab:gp-results-gl410}--~\ref{tab:gp-results-gl406} (see Sec.~\ref{sec:results-nights}).
Tables~\ref{tab:gl410-data}--~\ref{tab:gl699-data} present the $\langle B \rangle$ and d$Temp$ data used in this paper, and Tables~\ref{tab:gl410-data-bell}--~\ref{tab:gl699-data-bell} present the $B_\ell$  data used in this paper (first entries, full table available at CDS).

\begin{table}[h!]
	\renewcommand{\arraystretch}{1.125}
	\caption{GP hyper-parameters for DS~Leo (Gl~410).\label{tab:gp-results-gl410}}
	\begin{tabular}{lcc}
		\hline
		\hline 
		DS~Leo & Value & Prior \\ 
		\hline 
		\multicolumn{3}{c}{GP fit on $\langle B \rangle$}\\$\mu$~(kG) & $0.90^{+0.02}_{-0.02}$ & $\mathcal{U}(-10, 10)$ \\  
		$\sigma$~(kG) & $0.01^{+0.01}_{-0.01}$ & $\mathcal{U}(0, 1000)$ \\
		$\alpha$~(kG) & $0.08^{+0.01}_{-0.01}$ & $\mathcal{U}(0, 10)$ \\ 
		$l$~(d) & $54^{+8}_{-7}$ & $\mathcal{U}(10, 500)$ \\ 
		$\beta$ & $0.57^{+0.10}_{-0.08}$ & $\mathcal{U}(0, 20)$ \\ 
		$P_{\rm rot}$~(d) & $13.953^{+0.088}_{-0.086}$ & $\mathcal{U}(5, 20)$ \\ 
		$\chi^2_r$ &  0.72 &  \\
		RMS & $0.02$~(kG) &  \\
		\hline 
		\multicolumn{3}{c}{GP fit on $B_\ell$}\\$\mu_\ell$~(G) & $-0.60^{+3.24}_{-3.28}$ & $\mathcal{U}(-500, 500)$ \\ 
		$\sigma_\ell$~(G) & $1.94^{+0.94}_{-1.16}$ & $\mathcal{U}(0, 1000)$ \\ 
		$\alpha_\ell$~(G) & $15.19^{+1.92}_{-1.60}$ & $\mathcal{U}(0, 1000)$ \\ 
		$l_\ell$~(d) & $57^{+9}_{-7}$ & $\mathcal{U}(10, 500)$ \\ 
		$\beta_\ell$ & $0.45^{+0.06}_{-0.05}$ & $\mathcal{U}(0, 5)$ \\ 
		$P_{\rm rot,\ell}$~(d) & $14.001^{+0.077}_{-0.075}$ & $\mathcal{U}(5, 20)$ \\ 
		$\chi^2_{r,\ \ell}$ & 0.66 &  \\
		RMS$_{\ell}$ & $3.93$~(G) &  \\
		\hline 
		\multicolumn{3}{c}{GP fit on $\langle B \rangle$ and $B_\ell$}\\$\mu$~(kG) &  $0.90^{+0.02}_{-0.02}$ &  $\mathcal{U}(-10, 10)$ \\ 
		$\sigma$~(kG) &  $0.01^{+0.01}_{-0.01}$ &  $\mathcal{U}(0, 1000)$ \\ 
		$\alpha$~(kG) &  $0.08^{+0.01}_{-0.01}$ &  $\mathcal{U}(0, 10)$ \\ 
		$l$~(d) &  $54^{+8}_{-7}$ &  $\mathcal{U}(10, 500)$ \\ 
		$\beta$ &  $0.57^{+0.10}_{-0.08}$ &  $\mathcal{U}(0, 20)$ \\ 
		$P_{\rm rot}$~(d) &  $13.980^{+0.059}_{-0.058}$ &  $\mathcal{U}(5, 20)$ \\ 
		$\mu_\ell$~(G) &  $-0.76^{+3.31}_{-3.34}$ &  $\mathcal{U}(-500, 500)$ \\ 
		$\sigma_\ell$~(G) &  $1.90^{+0.97}_{-1.14}$ &  $\mathcal{U}(0, 1000)$ \\ 
		$\alpha_\ell$~(G) &  $15.48^{+1.98}_{-1.61}$ &  $\mathcal{U}(0, 1000)$ \\ 
		$l_\ell$~(d) &  $58^{+9}_{-7}$ &  $\mathcal{U}(10, 500)$ \\
		$\beta_\ell$ &  $0.45^{+0.06}_{-0.05}$ &  $\mathcal{U}(0, 20)$ \\ 
		$\chi^2_r$ &  0.72 &   \\
		$\chi^2_{r,\ \ell}$ &  0.66 &   \\
		RMS &  $0.02$~(kG) &   \\
		RMS$_{\ell}$ &  $3.94$~(G) &  \\
		\hline 
		\multicolumn{3}{c}{GP fit on d$Temp$}\\$\mu_{{\rm d}Temp}$~(K) & $-1.18^{+1.57}_{-1.61}$ & $\mathcal{U}(-50, 50)$ \\ 
		$\sigma_{{\rm d}Temp}$~(K) & $0.71^{+0.13}_{-0.12}$ & $\mathcal{U}(0, 1000)$ \\ 
		$\alpha_{{\rm d}Temp}$~(K) & $7.46^{+0.93}_{-0.81}$ & $\mathcal{U}(0, 100)$ \\ 
		$l_{{\rm d}Temp}$~(d) & $38^{+3}_{-3}$ & $\mathcal{U}(10, 500)$ \\ 
		$\beta_{{\rm d}Temp}$ & $0.63^{+0.07}_{-0.06}$ & $\mathcal{U}(0, 20)$ \\ 
		$P_{{\rm rot, d}Temp}$~(d) & $14.214^{+0.106}_{-0.102}$ & $\mathcal{U}(3, 200)$ \\ 
		$\chi^2_{r, {\rm d}Temp}$ & 0.98 &  \\
		RMS$_{{\rm d}Temp}$ & $0.58$~(K) &  \\
		\hline
	\end{tabular}
	\tablefoot{ Hyper-parameters obtained when applying our process to 
		the small-scale magnetic field measurements ($\langle B \rangle$), 
		the longitudinal magnetic field ($B_{\ell}$),
		or $\langle B \rangle$ and $B_{\ell}$ simultaneously. Error bars on $\langle B \rangle$ and d$Temp$ measurements were adjusted to ensure a minimum reduced $\chi^2$ close to one for the best fit.}
\end{table}
\begin{table}[h!]
	\renewcommand{\arraystretch}{1.125}
	\caption{Same as Table~\ref{tab:gp-results-gl410} for AD~Leo (Gl~388).\label{tab:gp-results-gl388}}
	\begin{tabular}{lcc}
		\hline
		\hline 
		AD~Leo & Value & Prior \\ 
		\hline 
		\multicolumn{3}{c}{GP fit on $\langle B \rangle$}\\$\mu$~(kG) &  $3.22^{+0.08}_{-0.07}$ &  $\mathcal{U}(-10, 10)$ \\ 
		$\sigma$~(kG) &  $0.01^{+0.01}_{-0.01}$ &  $\mathcal{U}(0, 1000)$ \\ 
		$\alpha$~(kG) &  $0.09^{+0.07}_{-0.04}$ &  $\mathcal{U}(0, 10)$ \\ 
		$l$~(d) &  $300$ &  Fixed \\ 
		$\beta$ &  $1.50$ &  Fixed \\ 
		$P_{\rm rot}$~(d) &  $2.232^{+0.095}_{-0.148}$ &  $\mathcal{G}(2.23, 0.20)$ \\ 
		$\chi^2_r$ &  0.74 &   \\
		RMS & $0.05$~(kG) &   \\
		\hline 
		\multicolumn{3}{c}{GP fit on $B_\ell$}\\$\mu_\ell$~(G) & $-164.94^{+118.82}_{-131.02}$ & $\mathcal{U}(-1000, 1000)$ \\ 
		$\sigma_\ell$~(G) & $5.02^{+3.50}_{-3.58}$ & $\mathcal{U}(0, 1000)$ \\ 
		$\alpha_\ell$~(G) & $79.87^{+114.76}_{-68.58}$ & $\mathcal{U}(0, 1000)$ \\ 
		$l_\ell$~(d) & $254^{+105}_{-93}$ & $\mathcal{U}(10, 500)$ \\ 
		$\beta_\ell$ & $1.42^{+1.32}_{-0.94}$ & $\mathcal{U}(0, 5)$ \\ 
		$P_{\rm rot,\ell}$~(d) & $2.230^{+0.001}_{-0.001}$ & $\mathcal{U}(2, 5)$ \\ 
		$\chi^2_{r,\ \ell}$ & 0.94 &  \\
		RMS$_{\ell}$ & $15.58$~(G) &  \\
		\hline 
		\multicolumn{3}{c}{GP fit on $\langle B \rangle$ and $B_\ell$}\\$\mu$~(kG) &  $3.22^{+0.08}_{-0.08}$ &  $\mathcal{U}(-10, 10)$ \\ 
		$\sigma$~(kG) &  $0.02^{+0.01}_{-0.01}$ &  $\mathcal{U}(0, 1000)$ \\ 
		$\alpha$~(kG) &  $0.09^{+0.07}_{-0.04}$ &  $\mathcal{U}(0, 10)$ \\ 
		$l$~(d) &  $300$ &  Fixed \\ 
		$\beta$ &  $1.50$ &  Fixed \\ 
		$P_{\rm rot}$~(d) &  $2.230^{+0.001}_{-0.001}$ &  $\mathcal{U}(1, 5)$ \\ 
		$\mu_\ell$~(G) &  $-157.84^{+134.94}_{-147.75}$ &  $\mathcal{U}(-1000, 1000)$ \\ 
		$\sigma_\ell$~(G) &  $5.74^{+3.49}_{-3.89}$ &  $\mathcal{U}(0, 1000)$ \\ 
		$\alpha_\ell$~(G) &  $98.77^{+157.32}_{-85.53}$ &  $\mathcal{U}(0, 1000)$ \\ 
		$l_\ell$~(d) &  $282^{+157}_{-111}$ &  $\mathcal{U}(10, 800)$ \\ 
		$\beta_\ell$ &  $1.67^{+1.45}_{-1.05}$ &  $\mathcal{U}(0, 5)$ \\ 
		$\chi^2_r$ &  0.75 &   \\
		$\chi^2_{r,\ \ell}$ &  0.81 &   \\
		RMS &  $0.05$~(kG) &   \\
		RMS$_{\ell}$ &  $14.75$~(G) &  \\
		\hline 
		\multicolumn{3}{c}{GP fit on d$Temp$}\\$\mu_{{\rm d}Temp}$~(K) & $0.21^{+1.17}_{-1.17}$ & $\mathcal{U}(-20, 20)$ \\ 
		$\sigma_{{\rm d}Temp}$~(K) & $0.49^{+0.20}_{-0.24}$ & $\mathcal{U}(0, 1000)$ \\ 
		$\alpha_{{\rm d}Temp}$~(K) & $2.60^{+0.68}_{-0.51}$ & $\mathcal{U}(0, 20)$ \\ 
		$l_{{\rm d}Temp}$~(d) & $82^{+24}_{-25}$ & $\mathcal{U}(10, 400)$ \\ 
		$\beta_{{\rm d}Temp}$ & $1.50$ & Fixed \\ 
		$P_{{\rm rot, d}Temp}$~(d) & $2.227^{+0.010}_{-0.015}$ & $\mathcal{G}(2.30, 0.20)$ \\ 
		$\chi^2_{r, {\rm d}Temp}$ & 1.01 &  \\
		RMS$_{{\rm d}Temp}$ & $0.89$~(K) &  \\
		\hline
	\end{tabular}
\end{table}

\begin{table}[h!]
	\renewcommand{\arraystretch}{1.125}
	\caption{Same as Table~\ref{tab:gp-results-gl410} for EV~Lac (Gl~873).}
	\label{tab:gp-results-gl873}
	\begin{tabular}{lcc}
		\hline
		\hline 
		EV~Lac & Value & Prior \\ 
		\hline 
		\multicolumn{3}{c}{GP fit on $\langle B \rangle$}\\$\mu$~(kG) & $4.42^{+0.19}_{-0.21}$ & $\mathcal{U}(0, 10)$ \\ 
		$\sigma$~(kG) & $0.02^{+0.01}_{-0.01}$ & $\mathcal{U}(0, 1000)$ \\ 
		$\alpha$~(kG) & $0.34^{+0.14}_{-0.09}$ & $\mathcal{U}(0, 10)$ \\ 
		$l$~(d) & $438^{+69}_{-73}$ & $\mathcal{U}(100, 800)$ \\ 
		$\beta$ & $0.60^{+0.13}_{-0.11}$ & $\mathcal{U}(0, 10)$ \\ 
		$P_{\rm rot}$~(d) & $4.362^{+0.001}_{-0.001}$ & $\mathcal{U}(0, 10)$ \\ 
		$\chi^2_r$ & 0.92 &  \\
		RMS & $0.05$~(kG) &  \\
		\hline 
		\multicolumn{3}{c}{GP fit on $B_\ell$}\\$\mu_\ell$~(G) & $-69.61^{+58.73}_{-58.77}$ & $\mathcal{U}(-500, 500)$ \\ 
		$\sigma_\ell$~(G) & $6.74^{+6.00}_{-5.67}$ & $\mathcal{U}(0, 1000)$ \\ 
		$\alpha_\ell$~(G) & $181.48^{+33.44}_{-26.57}$ & $\mathcal{U}(0, 1000)$ \\ 
		$l_\ell$~(d) & $127^{+15}_{-13}$ & $\mathcal{U}(10, 500)$ \\ 
		$\beta_\ell$ & $0.48^{+0.06}_{-0.06}$ & $\mathcal{U}(0, 3)$ \\ 
		$P_{\rm rot,\ell}$~(d) & $4.371^{+0.003}_{-0.003}$ & $\mathcal{U}(3, 10)$ \\ 
		$\chi^2_{r,\ \ell}$ & 0.69 &  \\
		RMS$_{\ell}$ & $26.39$~(G) &  \\
		\hline 
		\multicolumn{3}{c}{GP fit on $\langle B \rangle$ and $B_\ell$}\\$\mu$~(kG) & $4.42^{+0.20}_{-0.21}$ & $\mathcal{U}(0, 10)$ \\ 
		$\sigma$~(kG) & $0.01^{+0.01}_{-0.01}$ & $\mathcal{U}(0, 1000)$ \\ 
		$\alpha$~(kG) & $0.35^{+0.15}_{-0.10}$ & $\mathcal{U}(0, 1000)$ \\ 
		$l$~(d) & $440^{+81}_{-132}$ & $\mathcal{U}(10, 800)$ \\ 
		$\beta$ & $0.59^{+0.14}_{-0.11}$ & $\mathcal{U}(0, 10)$ \\ 
		$P_{\rm rot}$~(d) & $4.363^{+0.001}_{-0.001}$ & $\mathcal{U}(0, 10)$ \\ 
		$\mu_\ell$~(G) & $-81.99^{+70.28}_{-71.74}$ & $\mathcal{U}(-500, 500)$ \\ 
		$\sigma_\ell$~(G) & $9.37^{+5.49}_{-6.18}$ & $\mathcal{U}(0, 1000)$ \\ 
		$\alpha_\ell$~(G) & $191.74^{+40.29}_{-31.22}$ & $\mathcal{U}(0, 1000)$ \\ 
		$l_\ell$~(d) & $127^{+14}_{-13}$ & $\mathcal{U}(10, 800)$ \\ 
		$\beta_\ell$ & $0.54^{+0.07}_{-0.06}$ & $\mathcal{U}(0, 10)$ \\ 
		$\chi^2_r$ & 0.92 &  \\
		$\chi^2_{r,\ \ell}$ & 0.77 &  \\
		RMS & $0.05$~(kG) &  \\
		RMS$_{\ell}$ & $27.64$~(G) &  \\
		\hline 
		\multicolumn{3}{c}{GP fit on d$Temp$}\\$\mu_{{\rm d}Temp}$~(K) & $1.18^{+1.73}_{-1.44}$ & $\mathcal{U}(0, 10)$ \\ 
		$\sigma_{{\rm d}Temp}$~(K) & $0.53^{+0.21}_{-0.26}$ & $\mathcal{U}(0, 1000)$ \\ 
		$\alpha_{{\rm d}Temp}$~(K) & $6.25^{+1.61}_{-1.32}$ & $\mathcal{U}(0, 10)$ \\ 
		$l_{{\rm d}Temp}$~(d) & $460^{+92}_{-84}$ & $\mathcal{U}(100, 800)$ \\ 
		$\beta_{{\rm d}Temp}$ & $0.50^{+0.09}_{-0.07}$ & $\mathcal{U}(0, 10)$ \\ 
		$P_{{\rm rot, d}Temp}$~(d) & $4.362^{+0.001}_{-0.001}$ & $\mathcal{U}(0, 10)$ \\ 
		$\chi^2_{r, {\rm d}Temp}$ & 0.90 &  \\
		RMS$_{{\rm d}Temp}$ & $1.28$~(K) &  \\
		\hline
	\end{tabular}
\end{table}

\begin{table}[h!]
	\renewcommand{\arraystretch}{1.125}
	\caption{{Same as Table~\ref{tab:gp-results-gl410} for Barnard's star (Gl~699).}}
	\label{tab:gp-results-gl699}
	\begin{tabular}{lcc}
		\hline
		\hline 
		Barnard's~star & Value & Prior \\ 
		\hline 
		\multicolumn{3}{c}{GP fit on $\langle B \rangle$}\\$\mu$~(kG) & $0.46^{+0.05}_{-0.05}$ & $\mathcal{U}(-10, 10)$ \\ 
		$\sigma$~(kG) & $0.00^{+0.00}_{-0.00}$ & $\mathcal{U}(0, 1000)$ \\ 
		$\alpha$~(kG) & $0.10^{+0.05}_{-0.03}$ & $\mathcal{U}(0, 10)$ \\ 
		$l$~(d) & $222^{+58}_{-50}$ & $\mathcal{U}(50, 500)$ \\ 
		$\beta$ & $1.10^{+0.76}_{-0.44}$ & $\mathcal{U}(0, 20)$ \\ 
		$P_{\rm rot}$~(d) & $144^{+6}_{-6}$ & $\mathcal{U}(50, 200)$ \\ 
		$\chi^2_r$ & 0.91 &  \\
		RMS & $0.08$~(kG) &  \\
		\hline 
		\multicolumn{3}{c}{GP fit on $B_\ell$}\\$\mu_\ell$~(G) & $1.54^{+1.18}_{-1.17}$ & $\mathcal{U}(-500, 500)$ \\ 
		$\sigma_\ell$~(G) & $2.13^{+0.42}_{-0.48}$ & $\mathcal{U}(0, 1000)$ \\ 
		$\alpha_\ell$~(G) & $5.40^{+0.69}_{-0.63}$ & $\mathcal{U}(0, 1000)$ \\ 
		$l_\ell$~(d) & $100$ & Fixed \\ 
		$\beta_\ell$ & $0.40$ & Fixed \\ 
		$P_{\rm rot,\ell}$~(d) & $159.226^{+15.785}_{-19.423}$ & $\mathcal{U}(50, 200)$ \\ 
		$\chi^2_{r,\ \ell}$ & 1.09 &  \\
		RMS$_{\ell}$ & $5.00$~(G) &  \\
		\hline 
		\multicolumn{3}{c}{GP fit on $\langle B \rangle$ and $B_\ell$}\\$\mu$~(kG) & $0.47^{+0.03}_{-0.03}$ & $\mathcal{U}(-10, 10)$ \\ 
		$\sigma$~(kG) & $0.00^{+0.00}_{-0.00}$ & $\mathcal{U}(0, 1000)$ \\ 
		$\alpha$~(kG) & $0.09^{+0.02}_{-0.01}$ & $\mathcal{U}(0, 10)$ \\ 
		$l$~(d) & $181^{+32}_{-35}$ & $\mathcal{U}(10, 800)$ \\ 
		$\beta$ & $0.66^{+0.17}_{-0.12}$ & $\mathcal{U}(0, 20)$ \\ 
		$P_{\rm rot}$~(d) & $139^{+5}_{-4}$ & $\mathcal{U}(50, 200)$ \\ 
		$\mu_\ell$~(G) & $1.62^{+1.22}_{-1.18}$ & $\mathcal{U}(-1000, 1000)$ \\ 
		$\sigma_\ell$~(G) & $1.96^{+0.48}_{-0.57}$ & $\mathcal{U}(0, 1000)$ \\ 
		$\alpha_\ell$~(G) & $5.24^{+0.68}_{-0.59}$ & $\mathcal{U}(0, 1000)$ \\ 
		$l_\ell$~(d) & $100$ & Fixed \\ 
		$\beta_\ell$ & $0.40$ & Fixed \\ 
		$\chi^2_r$ & 0.81 &  \\
		$\chi^2_{r,\ \ell}$ & 1.08 &  \\
		RMS & $0.08$~(kG) &  \\
		RMS$_{\ell}$ & $4.98$~(G) &  \\
		\hline 
		\multicolumn{3}{c}{GP fit on d$Temp$}\\$\mu_{{\rm d}Temp}$~(K) & $-0.29^{+0.26}_{-0.26}$ & $\mathcal{U}(-50, 50)$ \\ 
		$\sigma_{{\rm d}Temp}$~(K) & $0.17^{+0.05}_{-0.07}$ & $\mathcal{U}(0, 1000)$ \\ 
		$\alpha_{{\rm d}Temp}$~(K) & $1.06^{+0.15}_{-0.12}$ & $\mathcal{U}(0, 100)$ \\ 
		$l_{{\rm d}Temp}$~(d) & $116^{+22}_{-24}$ & $\mathcal{U}(50, 500)$ \\ 
		$\beta_{{\rm d}Temp}$ & $0.49^{+0.12}_{-0.09}$ & $\mathcal{U}(0, 5)$ \\ 
		$P_{{\rm rot, d}Temp}$~(d) & $141.058^{+6.603}_{-11.625}$ & $\mathcal{U}(50, 200)$ \\ 
		$\chi^2_{r, {\rm d}Temp}$ & 0.99 &  \\
		RMS$_{{\rm d}Temp}$ & $0.47$~(K) &  \\
		\hline
	\end{tabular}
\end{table}

\begin{table}[h!]
	\renewcommand{\arraystretch}{1.125}
	\caption{{Same as Table~\ref{tab:gp-results-gl410} for PM~J18482+0741.}}
	\label{tab:gp-results-pmj}
	\begin{tabular}{lcc}
		\hline
		\hline 
		PM~J18482+0741 & Value & Prior \\ 
		\hline 
		\multicolumn{3}{c}{GP fit on $\langle B \rangle$}\\$\mu$~(kG) & $1.55^{+0.11}_{-0.11}$ & $\mathcal{U}(-10, 10)$ \\ 
		$\sigma$~(kG) & $0.01^{+0.02}_{-0.01}$ & $\mathcal{U}(0, 1000)$ \\ 
		$\alpha$~(kG) & $0.15^{+0.06}_{-0.05}$ & $\mathcal{U}(0, 10)$ \\ 
		$l$~(d) & $113^{+33}_{-31}$ & $\mathcal{U}(10, 500)$ \\ 
		$\beta$ & $1.50$ & Fixed \\ 
		$P_{\rm rot}$~(d) & $2.762^{+0.009}_{-0.007}$ & $\mathcal{G}(2.76, 0.60)$ \\ 
		$\chi^2_r$ & 0.89 &  \\
		RMS & $0.09$~(kG) &  \\
		\hline 
		\multicolumn{3}{c}{GP fit on $B_\ell$}\\$\mu_\ell$~(G) & $-6.20^{+92.85}_{-91.69}$ & $\mathcal{U}(-1000, 1000)$ \\ 
		$\sigma_\ell$~(G) & $8.40^{+3.30}_{-4.10}$ & $\mathcal{U}(0, 1000)$ \\ 
		$\alpha_\ell$~(G) & $44.18^{+126.91}_{-60.46}$ & $\mathcal{U}(0, 400)$ \\ 
		$l_\ell$~(d) & $650^{+177}_{-261}$ & $\mathcal{U}(10, 1000)$ \\ 
		$\beta_\ell$ & $0.35^{+1.30}_{-0.76}$ & $\mathcal{U}(0, 5)$ \\ 
		$P_{\rm rot,\ell}$~(d) & $2.761^{+0.001}_{-0.001}$ & $\mathcal{U}(2, 5)$ \\ 
		$\chi^2_{r,\ \ell}$ & 1.00 &  \\
		RMS$_{\ell}$ & $19.64$~(G) &  \\
		\hline 
		\multicolumn{3}{c}{GP fit on $\langle B \rangle$ and $B_\ell$}\\$\mu$~(kG) &  $1.55^{+0.31}_{-0.68}$ &  $\mathcal{U}(-10, 10)$ \\ 
		$\sigma$~(kG) &  $0.01^{+0.02}_{-0.01}$ &  $\mathcal{U}(0, 1000)$ \\ 
		$\alpha$~(kG) &  $0.15^{+1.40}_{-0.34}$ &  $\mathcal{U}(0, 5)$ \\ 
		$l$~(d) &  $112^{+100}_{-67}$ &  $\mathcal{U}(10, 500)$ \\ 
		$\beta$ &  $1.23^{+14.47}_{-3.43}$ &  $\mathcal{U}(0, 50)$ \\ 
		$P_{\rm rot}$~(d) &  $2.759^{+0.002}_{-0.002}$ &  $\mathcal{U}(2, 5)$ \\ 
		$\mu_\ell$~(G) &  $-12.02^{+56.52}_{-59.46}$ &  $\mathcal{U}(-1000, 1000)$ \\ 
		$\sigma_\ell$~(G) &  $11.47^{+3.06}_{-3.48}$ &  $\mathcal{U}(0, 1000)$ \\ 
		$\alpha_\ell$~(G) &  $81.00^{+24.94}_{-18.59}$ &  $\mathcal{U}(0, 400)$ \\ 
		$l_\ell$~(d) &  $300$ &  Fixed \\ 
		$\beta_\ell$ &  $1.00$ &  Fixed \\ 
		$\chi^2_r$ &  0.85 &   \\
		$\chi^2_{r,\ \ell}$ &  1.17 &   \\
		RMS &  $0.09$~(kG) &   \\
		RMS$_{\ell}$ &  $20.50$~(G) &   \\
		\hline 
		\multicolumn{3}{c}{GP fit on d$Temp$}\\$\mu_{{\rm d}Temp}$~(K) & $-2.36^{+3.54}_{-3.80}$ & $\mathcal{U}(-50, 50)$ \\ 
		$\sigma_{{\rm d}Temp}$~(K) & $0.37^{+0.56}_{-0.59}$ & $\mathcal{U}(0, 1000)$ \\ 
		$\alpha_{{\rm d}Temp}$~(K) & $4.11^{+3.11}_{-1.57}$ & $\mathcal{U}(0, 100)$ \\ 
		$l_{{\rm d}Temp}$~(d) & $300$ & Fixed \\ 
		$\beta_{{\rm d}Temp}$ & $1.50$ & Fixed \\ 
		$P_{{\rm rot, d}Temp}$~(d) & $2.771^{+1.426}_{-1.216}$ & $\mathcal{U}(2, 5)$ \\ 
		$\chi^2_{r, {\rm d}Temp}$ & 0.98 &  \\
		RMS$_{{\rm d}Temp}$ & $3.16$~(K) &  \\
		\hline
	\end{tabular}
\end{table}

\begin{table}[h!]
	\renewcommand{\arraystretch}{1.125}
	\caption{{Same as Table~\ref{tab:gp-results-gl410} for CN~Leo (Gl~406).}}
	\label{tab:gp-results-gl406}
	\begin{tabular}{lcc}
		\hline
		\hline 
		CN~Leo & Value & Prior \\ 
		\hline 
		\multicolumn{3}{c}{GP fit on $\langle B \rangle$}\\$\mu$~(kG)   $3.30^{+0.06}_{-0.06}$ & $\mathcal{U}(-10, 10)$ \\ 
		$\sigma$~(kG) & $0.01^{+0.01}_{-0.01}$ & $\mathcal{U}(0, 1000)$ \\ 
		$\alpha$~(kG) & $0.09^{+0.03}_{-0.02}$ & $\mathcal{U}(0, 10)$ \\ 
		$l$~(d) & $300$ & Fixed \\ 
		$\beta$ & $1.50$ & Fixed \\ 
		$P_{\rm rot}$~(d) & $2.694^*$ & $\mathcal{G}(2.70, 0.30)$ \\ 
		$\chi^2_r$ & 0.89 &  \\
		RMS & $0.15$~(kG) &  \\
		\hline 
		\multicolumn{3}{c}{GP fit on $B_\ell$}\\$\mu_\ell$~(G) & $-622.91^{+93.31}_{-104.48}$ & $\mathcal{U}(-1000, 1000)$ \\ 
		$\sigma_\ell$~(G) & $7.70^{+8.55}_{-8.05}$ & $\mathcal{U}(0, 1000)$ \\ 
		$\alpha_\ell$~(G) & $107.30^{+138.78}_{-76.88}$ & $\mathcal{U}(0, 1000)$ \\ 
		$l_\ell$~(d) & $99^{+35}_{-49}$ & $\mathcal{U}(10, 500)$ \\ 
		$\beta_\ell$ & $1.74^{+2.56}_{-1.36}$ & $\mathcal{U}(0, 10)$ \\ 
		$P_{\rm rot,\ell}$~(d) & $2.696^{+0.004}_{-0.006}$ & $\mathcal{U}(2, 4)$ \\ 
		$\chi^2_{r,\ \ell}$ & 0.88 &  \\
		RMS$_{\ell}$ & $53.53$~(G) &  \\
		\hline 
		\multicolumn{3}{c}{GP fit on $\langle B \rangle$ and $B_\ell$}\\$\mu$~(kG) & $3.31^{+0.19}_{-0.20}$ & $\mathcal{U}(-10, 10)$ \\ 
		$\sigma$~(kG) & $0.01^{+0.01}_{-0.01}$ & $\mathcal{U}(0, 1000)$ \\ 
		$\alpha$~(kG) & $0.17^{+0.22}_{-0.12}$ & $\mathcal{U}(0, 10)$ \\ 
		$l$~(d) & $498^{+263}_{-221}$ & $\mathcal{U}(10, 1000)$ \\ 
		$\beta$ & $7.56^{+5.85}_{-5.48}$ & $\mathcal{U}(0, 20)$ \\ 
		$P_{\rm rot}$~(d) & $2.700^{+0.002}_{-0.004}$ & $\mathcal{U}(2, 5)$ \\ 
		$\mu_\ell$~(G) & $-649.35^{+103.79}_{-116.08}$ & $\mathcal{U}(-1000, 1000)$ \\ 
		$\sigma_\ell$~(G) & $10.26^{+9.54}_{-8.98}$ & $\mathcal{U}(0, 1000)$ \\ 
		$\alpha_\ell$~(G) & $163.57^{+157.74}_{-88.07}$ & $\mathcal{U}(0, 1000)$ \\ 
		$l_\ell$~(d) & $151^{+34}_{-42}$ & $\mathcal{U}(10, 1000)$ \\ 
		$\beta_\ell$ & $2.78^{+2.64}_{-1.59}$ & $\mathcal{U}(0, 10)$ \\ 
		$\chi^2_r$ & 0.93 &  \\
		$\chi^2_{r,\ \ell}$ & 0.94 &  \\
		RMS & $0.15$~(kG) &  \\
		RMS$_{\ell}$ & $54.59$~(G) &  \\
		\hline 
		\multicolumn{3}{c}{GP fit on d$Temp$}\\$\mu_{{\rm d}Temp}$~(K) & $0.20^{+0.43}_{-0.43}$ & $\mathcal{U}(-1000, 1000)$ \\ 
		$\sigma_{{\rm d}Temp}$~(K) & $0.68^{+0.28}_{-0.35}$ & $\mathcal{U}(0, 1000)$ \\ 
		$\alpha_{{\rm d}Temp}$~(K) & $1.53^{+0.36}_{-0.29}$ & $\mathcal{U}(0, 1000)$ \\ 
		$l_{{\rm d}Temp}$~(d) & $45^{+15}_{-13}$ & $\mathcal{U}(10, 500)$ \\ 
		$\beta_{{\rm d}Temp}$ & $1.00$ & Fixed \\ 
		$P_{{\rm rot, d}Temp}$~(d) & $3.003^{+0.031}_{-0.536}$ & $\mathcal{G}(2.70, 0.30)$ \\ 
		$\chi^2_{r, {\rm d}Temp}$ & 1.02 &  \\
		RMS$_{{\rm d}Temp}$ & $1.78$~(K) &  \\
		\hline
	\end{tabular}
	\tablefoot{{$^*$ Value corresponding to the maximum of likelihood in the posterior distribution. The posterior distribution shows multiple peaks that does not allow us to extract statisfactory error bars.}}
\end{table}

\begin{table*}
	\caption{Measurements obtained each night for DS~Leo (abstract Table).\label{tab:gl410-data}} 
	\begin{tabular}{ccccccccc}
		\hline
		\hline
MJD & $\langle B \rangle$ & $f_0$ & $f_2$ & $f_4$ & $f_6$ & $f_8$ & $f_{10}$ & d$Temp$ \\
& (kG) & (\%) & (\%) & (\%) & (\%) & (\%) & (\%) & (K)\\
\hline
59157.6134 & $1.03\pm0.03$ & $51.6\pm1.5$ & $45.7\pm2.0$ & $2.4\pm1.2$ & $0.1\pm0.3$ & $0.1\pm0.2$ & $0.0\pm0.1$ & $-10.11\pm0.41$\\
59158.6077 & $1.08\pm0.03$ & $48.9\pm1.5$ & $48.4\pm1.9$ & $2.4\pm1.1$ & $0.2\pm0.4$ & $0.1\pm0.2$ & $0.0\pm0.1$ & $-13.46\pm0.41$\\
59207.5509 & $0.97\pm0.04$ & $54.5\pm2.0$ & $43.2\pm2.5$ & $1.6\pm1.2$ & $0.5\pm0.6$ & $0.1\pm0.3$ & $0.1\pm0.2$ & $3.54\pm0.49$\\
		\hline
	\end{tabular}
	\tablefoot{The complete table is available online in machine readable format.}
\end{table*}

\begin{table*}
	\caption{Same as Table~\ref{tab:gl410-data} for EV~Lac\label{tab:gl873-data}} 
	\begin{tabular}{ccccccccc}
		\hline
		\hline
MJD & $\langle B \rangle$ & $f_0$ & $f_2$ & $f_4$ & $f_6$ & $f_8$ & $f_{10}$ & d$Temp$ \\
& (kG) & (\%) & (\%) & (\%) & (\%) & (\%) & (\%) & (K)\\
\hline
58737.5820 & $4.63\pm0.05$ & $1.5\pm4.0$ & $28.8\pm3.0$ & $30.0\pm4.8$ & $20.1\pm5.4$ & $16.3\pm4.5$ & $3.3\pm2.4$ & $-4.45\pm1.53$\\
58744.3863 & $4.47\pm0.05$ & $1.8\pm4.1$ & $35.9\pm3.2$ & $25.2\pm4.8$ & $20.5\pm5.3$ & $7.6\pm4.7$ & $9.0\pm2.7$ & $-0.74\pm1.49$\\
58745.3476 & $4.46\pm0.06$ & $1.2\pm4.5$ & $31.8\pm3.2$ & $36.0\pm5.2$ & $12.4\pm6.0$ & $10.6\pm5.2$ & $7.9\pm2.9$ & $0.53\pm1.89$\\
		\hline
	\end{tabular}
\end{table*}
	
\begin{table*}
	\caption{Same as Table~\ref{tab:gl410-data} for AD~Leo\label{tab:gl388-data}} 
	\begin{tabular}{ccccccccc}
		\hline
		\hline
MJD & $\langle B \rangle$ & $f_0$ & $f_2$ & $f_4$ & $f_6$ & $f_8$ & $f_{10}$ & d$Temp$ \\
& (kG) & (\%) & (\%) & (\%) & (\%) & (\%) & (\%) & (K)\\
\hline
58528.5699 & $3.26\pm0.07$ & $0.9\pm1.7$ & $51.7\pm3.3$ & $33.4\pm4.3$ & $12.0\pm3.7$ & $1.6\pm1.9$ & $0.4\pm0.8$ & $-2.06\pm0.94$\\
58530.4672 & $3.40\pm0.06$ & $0.3\pm1.0$ & $49.4\pm2.8$ & $34.3\pm4.3$ & $12.4\pm4.0$ & $3.3\pm2.3$ & $0.4\pm0.8$ & $-3.09\pm0.91$\\
58531.6235 & $3.34\pm0.07$ & $1.2\pm2.0$ & $52.0\pm3.3$ & $32.6\pm4.1$ & $7.5\pm4.1$ & $6.2\pm2.7$ & $0.5\pm1.1$ & $1.20\pm0.94$\\
		\hline
	\end{tabular}
\end{table*}

\begin{table*}
	\caption{Same as Table~\ref{tab:gl410-data} for CN~Leo\label{tab:gl406-data}} 
	\begin{tabular}{ccccccccc}
		\hline
		\hline
MJD & $\langle B \rangle$ & $f_0$ & $f_2$ & $f_4$ & $f_6$ & $f_8$ & $f_{10}$ & d$Temp$ \\
& (kG) & (\%) & (\%) & (\%) & (\%) & (\%) & (\%) & (K)\\
\hline
58528.5993 & $3.51\pm0.08$ & $14.4\pm7.8$ & $46.8\pm12.0$ & $12.4\pm9.5$ & $11.2\pm8.7$ & $13.1\pm6.6$ & $2.0\pm2.2$ & $-0.76\pm2.27$\\
58530.5844 & $3.37\pm0.08$ & $15.2\pm8.1$ & $45.6\pm12.9$ & $14.5\pm10.7$ & $13.5\pm9.1$ & $9.5\pm6.1$ & $1.7\pm1.8$ & $-7.09\pm2.14$\\
58531.6429 & $5.01\pm0.48$ & $35.7\pm13.6$ & $13.2\pm13.5$ & $11.8\pm13.0$ & $13.0\pm13.8$ & $13.9\pm14.1$ & $12.4\pm13.6$ & $4.77\pm2.30$\\
		\hline
	\end{tabular}
\end{table*}

\begin{table*}
	\caption{Same as Table~\ref{tab:gl410-data} for PMJ~18482+0741\label{tab:pmj-data}} 
	\begin{tabular}{ccccccccc}
		\hline
		\hline
MJD & $\langle B \rangle$ & $f_0$ & $f_2$ & $f_4$ & $f_6$ & $f_8$ & $f_{10}$ & d$Temp$ \\
& (kG) & (\%) & (\%) & (\%) & (\%) & (\%) & (\%) & (K)\\
\hline
59296.5872 & $1.66\pm0.09$ & $34.5\pm3.0$ & $58.6\pm8.2$ & $3.3\pm3.2$ & $2.0\pm1.8$ & $1.0\pm1.1$ & $0.6\pm0.6$ & $-1.65\pm3.08$\\
59297.6055 & $1.60\pm0.08$ & $32.3\pm2.6$ & $63.0\pm7.8$ & $2.5\pm2.5$ & $1.2\pm1.3$ & $0.7\pm0.7$ & $0.4\pm0.5$ & $-5.63\pm2.78$\\
59304.6175 & $1.59\pm0.09$ & $36.9\pm3.0$ & $56.7\pm8.5$ & $3.1\pm3.0$ & $1.7\pm1.7$ & $1.1\pm1.1$ & $0.6\pm0.6$ & $-1.13\pm3.33$\\
		\hline
	\end{tabular}
\end{table*}

\begin{table*}
	\caption{Same as Table~\ref{tab:gl410-data} for Barnard's star\label{tab:gl699-data}} 
	\begin{tabular}{ccccccccc}
		\hline
		\hline
MJD & $\langle B \rangle$ & $f_0$ & $f_2$ & $f_4$ & $f_6$ & $f_8$ & $f_{10}$ & d$Temp$ \\
 & (kG) & (\%) & (\%) & (\%) & (\%) & (\%) & (\%) & (K)\\
\hline
58528.6717 & $0.62\pm0.04$ & $68.9\pm3.9$ & $31.1\pm3.9$ & $0.0\pm0.0$ & $0.0\pm0.0$ & $0.0\pm0.0$ & $0.0\pm0.0$ & $-1.34\pm0.50$\\
58530.6687 & $0.55\pm0.05$ & $72.5\pm4.0$ & $27.5\pm4.0$ & $0.0\pm0.0$ & $0.0\pm0.0$ & $0.0\pm0.0$ & $0.0\pm0.0$ & $-1.85\pm0.24$\\
58540.6805 & $0.56\pm0.04$ & $71.9\pm3.9$ & $28.1\pm3.9$ & $0.0\pm0.0$ & $0.0\pm0.0$ & $0.0\pm0.0$ & $0.0\pm0.0$ & $-2.64\pm0.20$\\
		\hline
	\end{tabular}
\end{table*}

\begin{table}
	\caption{$B_\ell$ measurements obtained each night for DS~Leo (abstract Table). The complete table is available at CDS.\label{tab:gl410-data-bell}} 
	\begin{tabular}{cc}
		\hline
		\hline
		MJD & $B_\ell$ \\
		& (G) \\
		\hline
		59157.6116 & $-10.66\pm4.82$ \\
		59158.6059 & $1.14\pm4.72$ \\
		59207.5537 & $-3.18\pm4.77$ \\
		\hline
	\end{tabular}
\end{table}

\begin{table}
	\caption{Same as Table~\ref{tab:gl410-data} for EV~Lac\label{tab:gl873-data-bell}} 
	\begin{tabular}{cc}
		\hline
		\hline
		MJD & $B_\ell$ \\
			& (G) \\
		\hline
		58737.5865 & $128.63\pm33.70$ \\
		58744.3908 & $-186.34\pm32.31$ \\
		58745.3521 & $62.20\pm34.57$ \\
		\hline
	\end{tabular}
\end{table}

\begin{table}
	\caption{Same as Table~\ref{tab:gl410-data} for AD~Leo\label{tab:gl388-data-bell}} 
	\begin{tabular}{cc}
		\hline
		\hline
		MJD & $B_\ell$ \\
			& (G) \\
		\hline
		58588.2634 & $-223.42\pm17.46$ \\
		58589.5079 & $-199.36\pm20.36$ \\
		58591.3814 & $-217.51\pm17.10$ \\
		\hline
	\end{tabular}
\end{table}

\begin{table}
	\caption{Same as Table~\ref{tab:gl410-data} for CN~Leo\label{tab:gl406-data-bell}} 
	\begin{tabular}{cc}
		\hline
		\hline
		MJD & $B_\ell$ \\
			& (G) \\
		\hline
		58589.5176 & $-739.23\pm87.44$ \\
		58591.3895 & $-851.25\pm68.78$ \\
		58592.4557 & $-623.14\pm74.71$ \\
		\hline
	\end{tabular}
\end{table}

\begin{table}
	\caption{Same as Table~\ref{tab:gl410-data} for PMJ~18482+0741\label{tab:pmj18482+0741-data-bell}} 
	\begin{tabular}{cc}
		\hline
		\hline
		MJD & $B_\ell$ \\
			& (G) \\
		\hline
		59296.5882 & $-82.98\pm21.45$ \\
		59297.6060 & $14.88\pm18.11$ \\
		59304.6186 & $-42.17\pm26.33$ \\
		\hline
	\end{tabular}
\end{table}

\begin{table}
	\caption{Same as Table~\ref{tab:gl410-data} for Barnard's star\label{tab:gl699-data-bell}} 
	\begin{tabular}{cc}
		\hline
		\hline
		MJD & $B_\ell$ \\
			& (G) \\
		\hline
		58588.6111 & $-6.13\pm5.91$ \\
		58588.6163 & $2.76\pm5.97$ \\
		58589.6243 & $-0.69\pm6.22$ \\
		\hline
	\end{tabular}
\end{table}

\FloatBarrier
\section{filling factors distribution}
\begin{figure}[h!]
	\includegraphics[width=0.95\linewidth]{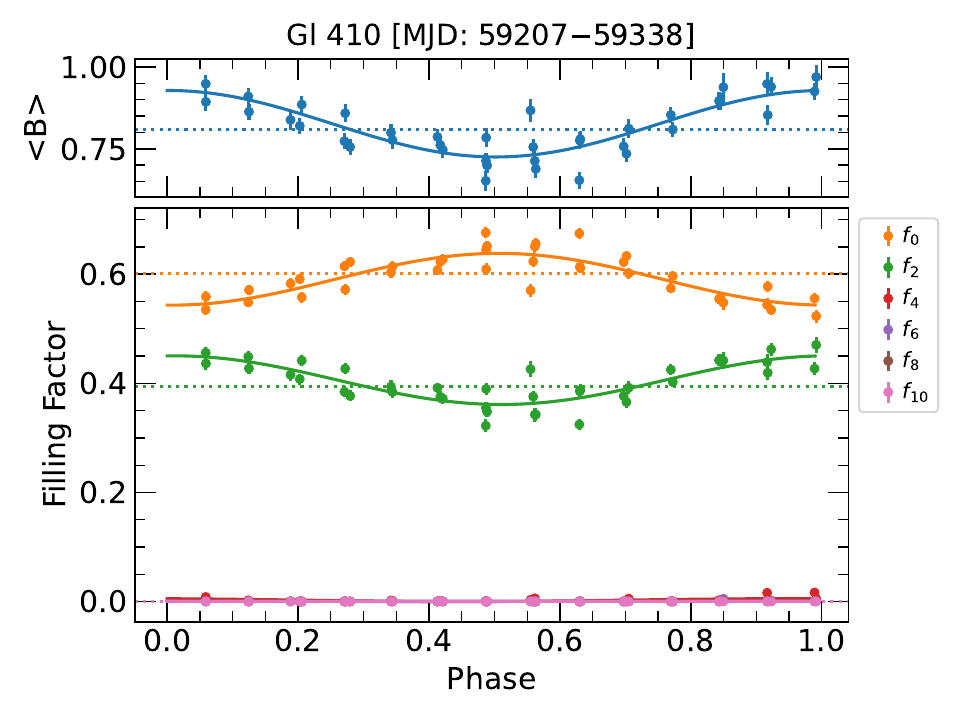}
	\caption{Same as Fig.~\ref{fig:b-distrib-gl873} for DS~Leo, assuming $P_{\rm rot}=13.982$\,d, for observations recorded between MJD 59207 and 59338.}
	\label{fig:b-distrib-gl410}
\end{figure}

\begin{figure}[h!]
	\includegraphics[width=0.95\linewidth]{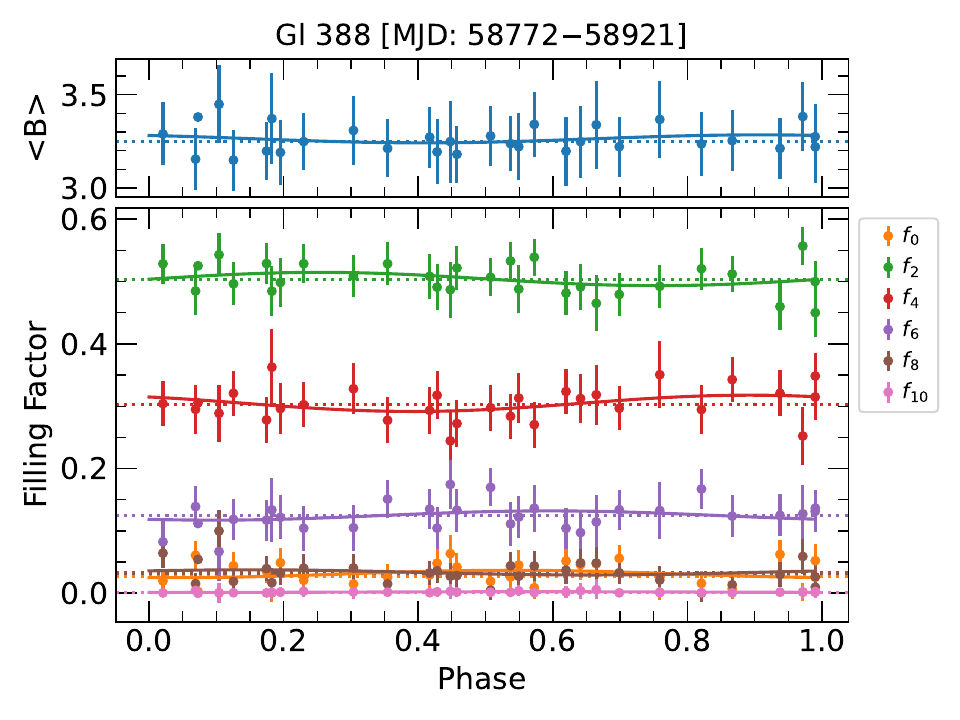}
	\caption{Same as Fig.~\ref{fig:b-distrib-gl873} for AD~Leo, assuming $P_{\rm rot}=2.230$\,d, for observations recorded between MJD 58772 and 58921.}
	\label{fig:b-distrib-gl388}
\end{figure}

\begin{figure}[h!]
	\includegraphics[width=0.95\linewidth]{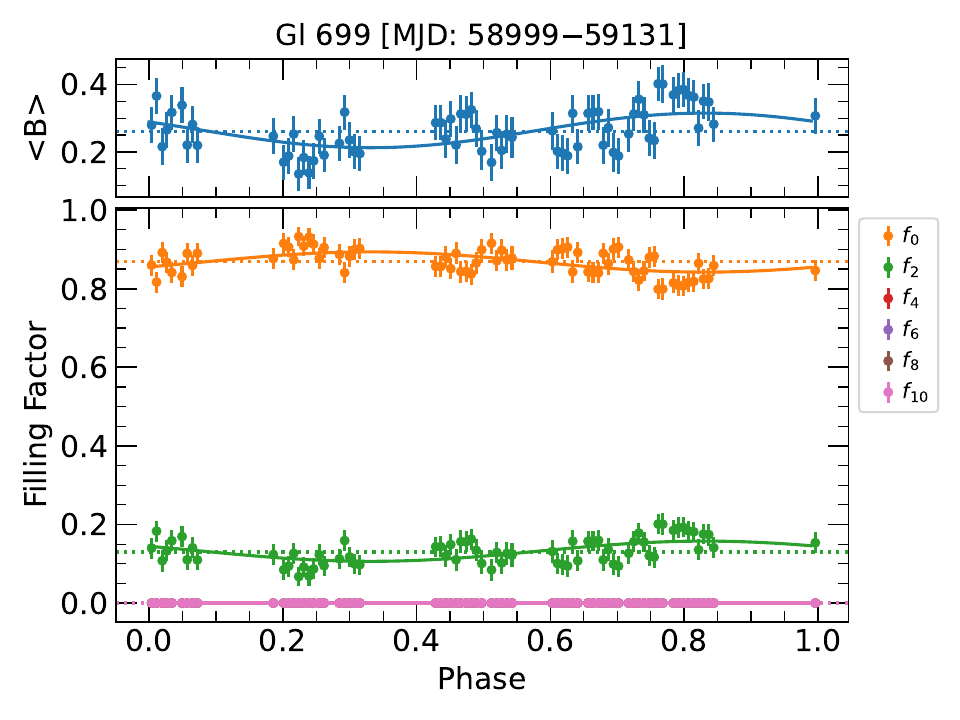}
	\caption{Same as Fig.~\ref{fig:b-distrib-gl873} for Barnard's star, assuming $P_{\rm rot}=131.472$\,d, for observations recorded between MJD 58999 and 59131.}
	\label{fig:b-distrib-gl699}
\end{figure}

\begin{figure}[h!]
	\includegraphics[width=0.95\linewidth]{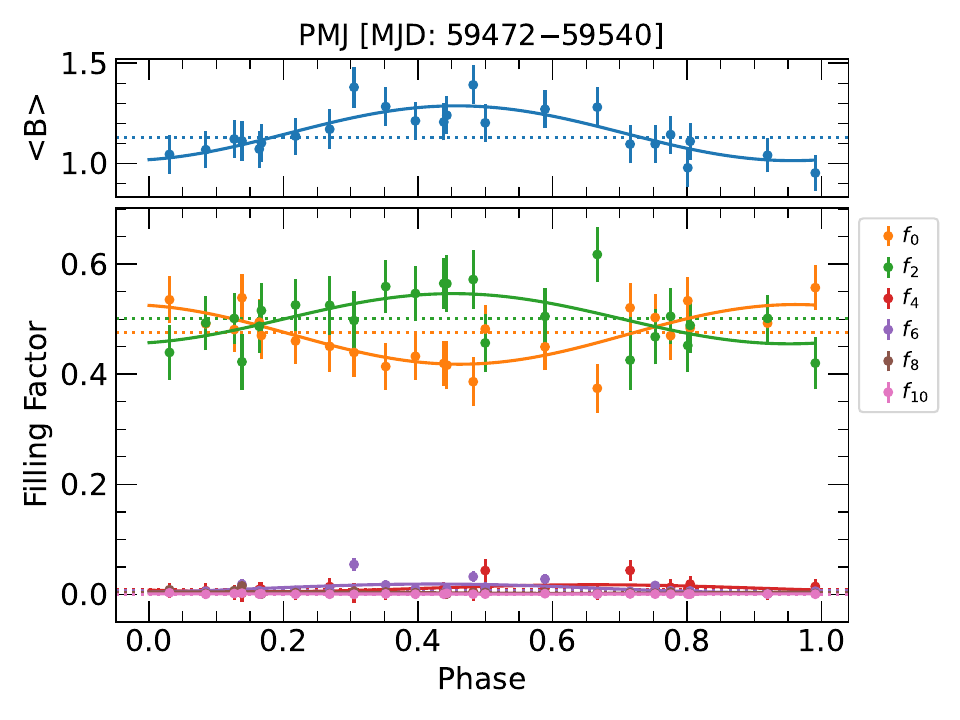}
	\caption{Same as Fig.~\ref{fig:b-distrib-gl873} for PMJ~J18482+0741, assuming ${P_{\rm rot}=2.760}$\,d, for observations recorded between MJD 59472 and 59540.}
	\label{fig:b-distrib-pmj}
\end{figure}

\begin{figure}[h!] 
	\includegraphics[width=0.95\linewidth]{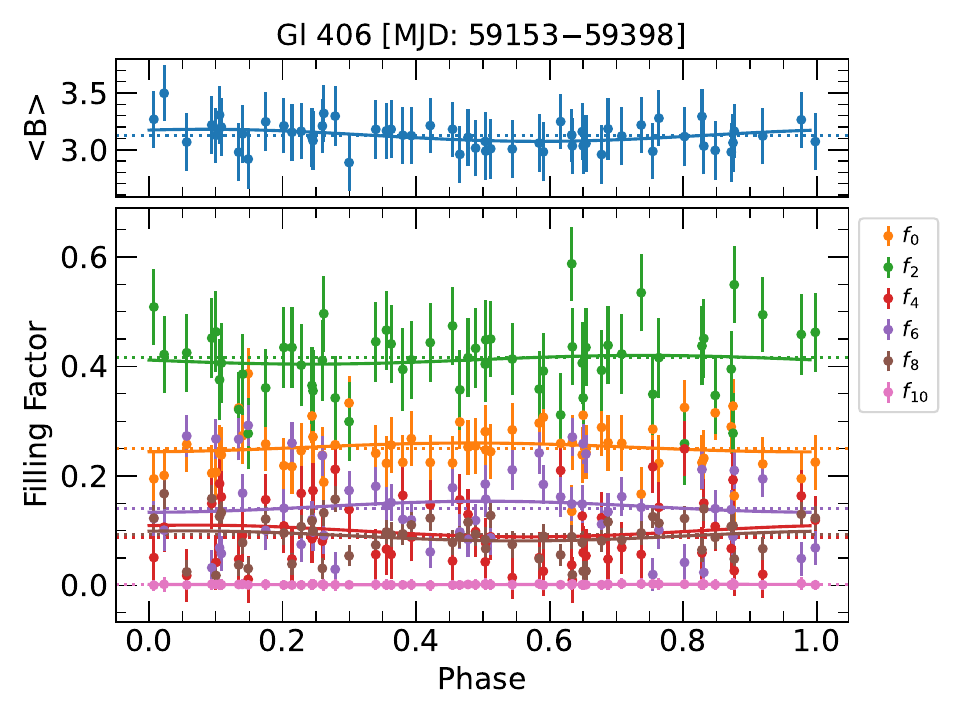}
	\caption{Same as Fig.~\ref{fig:b-distrib-gl873} for CN~Leo, assuming $P_{\rm rot}=2.696$\,d, for observations recorded between MJD 59153 and 59398.}
	\label{fig:b-distrib-gl406}
\end{figure}

Figures~\ref{fig:b-distrib-gl410},~\ref{fig:b-distrib-gl388},~\ref{fig:b-distrib-gl699},~\ref{fig:b-distrib-pmj}, and~\ref{fig:b-distrib-gl406} present the phase folded modulation of the magnetic filling factors DS~Leo, AD~Leo, Barnard's star, PMJ~J18482+0741 and CN~Leo, respectively.

\end{appendix}

\end{document}